\documentclass[12pt,hidelinks]{article}
\usepackage[body={17.5cm, 22cm},right=2cm]{geometry}
\usepackage{color}
\usepackage{amssymb,graphicx}
\usepackage{amsmath}
\usepackage{epsfig}
\usepackage[bookmarks=true,pdfborder={0 0 0}]{hyperref} 
\usepackage{cite}

\allowdisplaybreaks

\begin{document}
\setcounter{page}{0}
\thispagestyle{empty}

\parskip 3pt

\font\mini=cmr10 at 2pt

\begin{titlepage}
~\vspace{1cm}
\begin{center}

{\LARGE \bf The QCD axion, precisely}

\vspace{0.6cm}

{\large
Giovanni~Grilli~di~Cortona$^a$, Edward~Hardy$^b$, \\[6pt]
Javier~Pardo~Vega$^{a,b}$ and Giovanni~Villadoro$^{b}$}
\\
\vspace{.6cm}
{\normalsize { \sl $^{a}$ 
SISSA International School for Advanced Studies and INFN Trieste, \\
Via Bonomea 265, 34136, Trieste, Italy }}

\vspace{.3cm}
{\normalsize { \sl $^{b}$ Abdus Salam International Centre for Theoretical Physics, \\
Strada Costiera 11, 34151, Trieste, Italy}}

\end{center}
\vspace{.8cm}
\begin{abstract}
We show how several properties of the QCD axion can be extracted at high precision
using only first principle QCD computations. By combining NLO results
obtained in chiral perturbation theory with recent Lattice QCD results the full
axion potential, its mass and the coupling to photons can be reconstructed
with percent precision. Axion couplings to nucleons can also be derived reliably, 
with uncertainties smaller than ten percent. The approach presented here allows 
the precision to be further improved as uncertainties on the light quark masses
and the effective theory couplings are reduced.
We also compute the finite temperature dependence of the axion potential and 
its mass up to the crossover region. For higher temperature we point out the unreliability 
of the conventional instanton approach and study its impact on the computation of 
the axion relic abundance.
\end{abstract}

\end{titlepage}

\tableofcontents

\section{Introduction}
In the Standard Model, the sum of the QCD topological angle and the common quark mass phase, 
$\theta=\theta_0+{\rm arg~det}M_q$, is experimentally bounded to lie below ${\cal O}$(10$^{-10}$)
from the non-observation of the neutron electric dipole moment (EDM) \cite{Crewther:1979pi,Afach:2015sja}. 
While $\theta={\cal O}(1)$ 
would completely change the physics of nuclei, its effects rapidly decouple for smaller values,
already becoming irrelevant for $\theta \lesssim 10^{-1}\div 10^{-2}$. Therefore, its extremely small value 
does not seem to be necessary to explain any known large-distance physics. 
This, together with the 
fact that other phases in the Yukawa matrices are ${\cal O}(1)$ and that $\theta$ 
can receive non-decoupling contributions from CP-violating new physics at arbitrarily 
high scales, begs for a dynamical explanation of its tiny value. 

Among the known solutions, the QCD axion \cite{Peccei:1977hh,Wilczek:1977pj,Weinberg:1977ma,Kim:1979if,Shifman:1979if,
Zhitnitsky:1980tq,Dine:1981rt} is probably 
the most simple and robust: the SM is augmented with an extra pseudo-goldstone boson,
whose only non-derivative coupling is to the QCD topological charge and suppressed by 
the scale $f_a$. Such a coupling allows the effects of $\theta$ to be redefined away via a
shift of the axion field, whose vacuum expectation value (VEV) is then guaranteed to vanish~\cite{Vafa:1984xg}.
 It also produces a mass for the axion ${\cal O}(m_\pi f_\pi/f_a)$.
Extra model dependent derivative couplings may be present but they
do not affect the solution of the strong-CP problem.
Both the mass and the couplings of the QCD axion are thus controlled by a single scale $f_a$.

Presently astrophysical constraints bound $f_a$ between few~10$^8$~GeV (see for e.g. \cite{Raffelt:2006cw}) and few~$10^{17}$~GeV \cite{Arvanitaki:2009fg,Arvanitaki:2010sy,Arvanitaki:2014wva}. 
It has been known for a long time \cite{Preskill:1982cy,Abbott:1982af,Dine:1982ah} 
that in most of the available parameter space the axion
may explain the observed dark matter of the universe. Indeed, non-thermal production
from the misalignment mechanism can easily generate a suitable abundance of cold axions
for values of $f_a$ large enough, compatible with those allowed by current bounds. Such a feature
is quite model independent and, if confirmed, may give non-trivial constraints on early cosmology.

Finally axion-like particles seem to be a generic feature of string compactification.
The simplicity and robustness of the axion solution to the strong-CP problem, 
the fact that it could easily explain the dark matter abundance of our Universe 
and the way it naturally fits within string theory make it one of the best 
motivated particle beyond the Standard Model.

Because of the extremely small couplings allowed by astrophysical bounds, the quest to
discover the QCD axion is a very challenging endeavor. 
The ADMX experiment \cite{Asztalos:2009yp}
is expected to become sensitive to a new region of parameter space 
unconstrained by indirect searches soon. Other experiments are also being planned
and several new ideas have recently been proposed to directly probe the QCD axion
\cite{Armengaud:2014gea,Horns:2012jf,Budker:2013hfa,Arvanitaki:2014dfa}.
To enhance the tiny signal some of these experiments, including ADMX, 
exploit resonance effects  and the fact that, if the axion is dark matter, the line width
of the resonance is suppressed by $v^2 \sim 10^{-6}$ ($v$ being the virial velocity in our galaxy) 
\cite{Sikivie:1983ip,Krauss:1985ub}.
Should the axion be discovered by such experiments, its mass would be known with
a comparably high precision, ${\cal O}(10^{-6})$. Depending on the experiment
different axion couplings may also be extracted with a different accuracy.

Can we exploit such high precision in the axion mass and maybe couplings?
What can we learn from such measurements? 
Will we be able to infer the UV completion of the axion? and its cosmology?

In this paper we try to make a small step towards answering some of these questions.
Naively, high precision in QCD axion physics seems hopeless. After all most of its properties,
such as its mass, couplings to matter and relic abundance are dominated by non perturbative
QCD dynamics. On the contrary, we will show that high precision is within reach.
Given its extremely light mass, QCD chiral Lagrangians \cite{Weinberg:1978kz,Gasser:1983yg,Gasser:1984gg} can  be used reliably. 
Performing a NLO computation we are able to extract the axion mass,
self coupling and its full potential at the percent level. The coupling to photons can
be extracted with similar precision, as well as the tension of domain walls.
As a spin-off we provide estimates of the topological susceptibility and the 
quartic moment with similar precision and new estimates of some low energy constants.

We also describe a new strategy to extract the couplings to nucleons directly from first
principle QCD. At the moment the precision is not yet at the percent level, but there is
room for improvement as more lattice QCD results become available.

The computation of the axion potential can easily be extended to finite temperature.
In particular, at temperatures below the crossover ($T_c\sim$170~MeV) 
chiral Lagrangians allow the temperature dependence of the axion potential and its mass to be computed. Around $T_c$ there is no known reliable perturbative expansion under control
and non-perturbative methods, such as lattice QCD  \cite{Buchoff:2013nra,Trunin:2015yda},
are required. 

At higher temperatures, when QCD turns perturbative, 
one may be tempted to use the dilute instanton gas approximation, which is
expected to hold at large enough temperatures. We point out however that the bad  convergence of the perturbative QCD expansion at finite temperatures makes the standard instanton result completely unreliable for temperatures below $10^6\,{\rm GeV}$, explaining the large discrepancy observed in recent lattice QCD simulations  \cite{Berkowitz:2015aua,Borsanyi:2015cka}. We conclude with a study of the impact of such uncertainty in the computation of the axion relic abundance, providing updated plots
for the allowed axion parameter space.

For convenience we report the main numerical results of the paper here, for the mass
$$
m_a  =  5.70(6)(4)\, \mu{\rm eV}\, \left(\frac{10^{12} {\rm GeV}}{f_a} \right) \,,
$$ 
the coupling to photons
$$
g_{a\gamma \gamma} =  \frac{\alpha_{em}}{2\pi f_a} \left[ \frac{E}{N} -1.92(4)\right] \,,
$$
the couplings to nucleons  (for the hadronic KSVZ model for definiteness)
$$
c_p^{\rm KSVZ} =-0.47(3)\,,  \qquad c_n^{\rm KSVZ} = -0.02(3)\,,
$$
and for the self quartic coupling and the tension of the domain wall respectively
$$\lambda_a  =-0.346(22)\cdot \frac{m_a^2}{f_a^2}\,,  \qquad
\sigma_a  = 8.97(5) \, m_a f_a^2\,,
$$
where for the axion mass the first error is from the uncertainties of quark masses while the second is from higher order corrections. 
As a by-product we also provide  a high precision estimate of the topological susceptibility
and the quartic moment
$$
\chi_{top}^{1/4}=75.5(5)~{\rm MeV}\,, \qquad b_2=-0.029(2)\,. 
$$
More complete results, explicit analytic formulae and details about conventions can be found in the text. 
The impact on the axion abundance computation from different finite temperature behaviors
of the axion mass is shown in figs.~\ref{fig:fachi} and \ref{fig:fahi}. 

The rest of the paper is organized as follows. In section~\ref{sec:Te0} we first  briefly review
known leading order results for the axion properties and then present our new computations 
and numerical estimates for the various properties at zero temperature. 
In section~\ref{sec:Tne0} we give results for the temperature dependence of the axion
mass and potential at increasing temperatures and the implications for the axion dark matter
abundance. We summarize our conclusions in section~\ref{sec:concl}. Finally,
we provide the details about the input parameters used and report extra formulae 
in the appendices.

\section{The cool axion: $T=0$ properties} \label{sec:Te0}
At energies below the Peccei Quinn (PQ) and the electroweak (EW) breaking scales the axion dependent part of the Lagrangian, at leading order in $1/f_a$ and the weak couplings can be written, without loss of generality, as
\begin{equation} \label{eq:Laga}
{\cal L}_a=\frac12 (\partial_\mu a)^2+\frac{a}{f_a} \frac{\alpha_s}{8\pi}G_{\mu\nu}\tilde G^{\mu\nu}
+\frac14 \,a\, g_{a\gamma \gamma}^0 F_{\mu\nu}\tilde F^{\mu\nu}+\frac{\partial_\mu a }{2f_a} \, j^{\mu}_{a,0} \ ,
\end{equation}
where the second term defines $f_a$, the dual gluon field strength $\tilde G_{\mu\nu}=\frac12 \epsilon_{\mu\nu\rho\sigma}G^{\rho\sigma}$, color indices are implicit, 
and the coupling to the photon field strength $F_{\mu\nu}$ is
\begin{equation}
g_{a\gamma \gamma}^0=\frac{\alpha_{em}}{2\pi f_a} \frac{E}{N}\,,
\end{equation}
where $E/N$ is the ratio of the Electromagnetic (EM) and the color anomaly (=8/3 for complete SU(5) representations).
Finally in the last term of eq.~(\ref{eq:Laga}) $j_{a,0}^\mu=c_q^0 \bar q \gamma^\mu \gamma_5 q$ 
is a model dependent axial current made of SM matter fields. The axionic pseudo shift-symmetry,
$a\to a+\delta$, has been used to remove the QCD $\theta$ angle.

The only non-derivative coupling to QCD can be conveniently reshuffled by a quark field redefinition. 
In particular performing a change of field variables on the up and down quarks
\begin{equation} \label{eq:redq}
q= \left(\begin{array}{c} u \\ d \end{array}\right) \to
e^{i \gamma_5 \frac{a}{2f_a} Q_a }\left(\begin{array}{c} u \\ d \end{array}\right)\,,\qquad {\rm tr}\,Q_a=1\,,
\end{equation}
eq.~(\ref{eq:Laga})  becomes
\begin{equation} \label{eq:Laga2}
{\cal L}_a=\frac12 (\partial_\mu a)^2 
+\frac14 \,a\, g_{a\gamma \gamma} F_{\mu\nu}\tilde F^{\mu\nu}+\frac{\partial_\mu a }{2 f_a}\,j^{\mu}_a 
- \bar q_L M_a q_R + h.c. \ ,
\end{equation}
where
\begin{align}\label{eq:Laga2def}
 g_{a\gamma \gamma}  = \frac{\alpha_{em}}{2\pi f_a} \left[\frac{E}{N} - 6\,{\rm tr} \left ( Q_a Q^2 \right )\right]\,, & \qquad j_a^\mu  = j_{a,0}^\mu - \bar q \gamma^\mu \gamma_5 Q_a q \,,  \\ & \nonumber \\
M_a=e^{i  \frac{a}{2 f_a} Q_a }M_q\, e^{i  \frac{a}{2 f_a} Q_a } \,, 
\qquad M_q=&\left( \begin{array}{c c}  m_u & 0 \\ 0 & m_d \end{array} \right ) \,,  \qquad  Q  =\left( \begin{array}{c c}  \textstyle \frac23 & 0 \\ 0 & -  \textstyle \frac13 \end{array} \right )\,.  \nonumber
\end{align}

The advantage of this basis of axion couplings is twofold. First the axion coupling to the axial current only renormalizes multiplicatively unlike the coupling to the gluon operator, which mixes with the axial current
divergence at one-loop. Second the only non-derivative couplings of the axion appear through the quark mass terms. 

At leading order in $1/f_a$ the axion can be treated as an external source, the effects from virtual axions being further suppressed by the tiny coupling. The non derivative couplings to QCD are encoded in the phase dependence of the dressed quark mass matrix $M_a$, while in the derivative couplings the axion enters as an external axial current.
The low energy behaviour of correlators involving such external sources is completely captured by chiral Lagrangians, whose \emph{raison d'\^etre} is exactly to provide a consistent perturbative expansion for such quantities. 

Notice that the choice of field redefinition (\ref{eq:redq}) allowed us to move the non-derivative 
couplings entirely into the lightest two quarks. In this way we can integrate out all the other quarks and  directly work in the 2-flavor effective theory, with $M_a$ capturing the whole axion dependence, at least for observables that do not depend on the derivative couplings.

At the leading order in the chiral expansion all the non-derivative dependence on the axion
is thus contained in the pion mass terms:
\begin{equation} \label{eq:chpt-mass}
{\cal L}_{p^2} \supset 2 B_0 \frac{f_\pi^2}{4} \langle U M_a^\dagger + M_a U^\dagger \rangle \,,
\end{equation}
where
\begin{equation}
U=e^{i\Pi/f_\pi}\,, \quad \Pi=\left( \begin{array}{cc} \pi^0 & \sqrt2 \pi^+ \\ 
\sqrt2 \pi^- & -\pi^0 \end{array} \right)\,,
\end{equation}
$\langle\cdots \rangle$ is the trace over flavor indices, $B_0$ is related to the chiral condensate and determined by the pion mass in term of the quark masses, and the pion decay constant is normalized such that $f_\pi\simeq 92$~MeV.

In order to derive the leading order effective axion potential we need only consider the neutral pion sector.
Choosing $Q_a$ proportional to the identity we have
\begin{align} \label{eq:potap}
V(a,\pi^0) & = - B_0 f_\pi^2 \left [m_u \cos \left(\textstyle \frac{\pi^0}{f_\pi}- \frac{a}{2f_a}\right)
+m_d \cos \left(\textstyle \frac{\pi^0}{f_\pi}+\frac{a}{2f_a}\right) \right ] \nonumber \\
& = - m_\pi^2 f_\pi^2 \sqrt{1-\frac{4 m_u m_d}{(m_u+m_d)^2}\sin^2 \left(\frac{a}{2f_a}\right) }\ \cos \left(\frac{\pi^0}{f_\pi}-\phi_a\right)
\end{align}
where 
\begin{equation}
\tan \phi_a\equiv \frac{m_u-m_d}{m_d+m_u} \tan\left (\frac{a}{2 f_a} \right)\,.
\end{equation}
On the vacuum $\pi^0$ gets a vacuum expectation value (VEV) proportional to $\phi_a$ to minimize the potential, the last cosine in eq.~(\ref{eq:potap}) is 1 on the vacuum, and $\pi^0$  can be trivially integrated out leaving the  axion effective potential 
\begin{equation} \label{eq:pota}
V(a)=-m_\pi^2 f_\pi^2 \sqrt{1-\frac{4 m_u m_d}{(m_u+m_d)^2}\sin^2 \left(\frac{a}{2f_a}\right) }\,.
\end{equation}
As expected the minimum is at $\langle a \rangle=0$ (thus solving the strong CP problem). Expanding to quadratic order we get the well-known \cite{Weinberg:1977ma} formula for the axion mass
\begin{equation}
m_a^2=\frac{m_u m_d}{(m_u+m_d)^2}\frac{m_\pi^2 f_\pi^2}{f_a^2} \,.
\end{equation}

Although the expression for the potential (\ref{eq:pota}) was derived long ago \cite{DiVecchia:1980ve},  
we would like to stress some points often under-emphasized in the literature.

\begin{figure}
\centering
\includegraphics[scale=0.3]{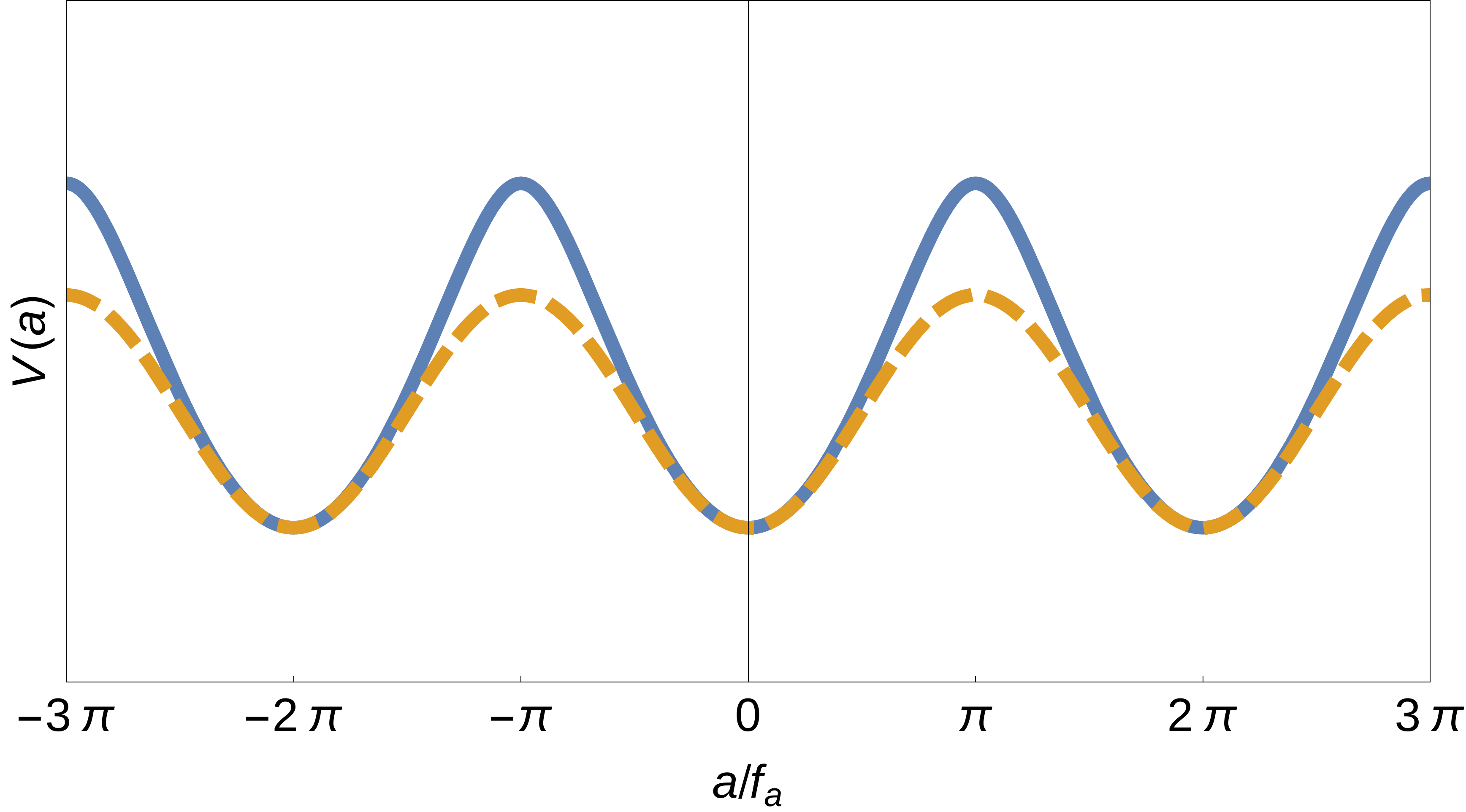}
\caption{\label{fig:pots} Comparison between the axion potential predicted by chiral Lagrangians, eq.~(\ref{eq:pota}) (continuous line) and the single cosine instanton one, $V^{inst}(a)=-m_a^2 f_a^2 \cos(a/f_a)$ (dashed line).}
\end{figure}

The axion potential (\ref{eq:pota}) is nowhere close to the single cosine suggested by the instanton calculation (see fig.~\ref{fig:pots}). 
This is not surprising given that the latter relies on a semiclassical approximation, which 
 is not under control in this regime. Indeed 
the shape of the potential is ${\cal O}(1)$ different from that of a single cosine,
and its dependence on the quark masses is non-analytic, as a consequence of the presence of light Goldstone modes. 
The axion self coupling, which is extracted from the fourth derivative of the potential
\begin{equation}
\lambda_a \equiv \left. \frac{\partial^4 V(a)}{\partial a^4} \right|_{a=0}=-\frac{m_u^2-m_u m_d+m_d^2}{(m_u+m_d)^2}\,\frac{m_a^2}{f_a^2}\,,
\end{equation}
is roughly a factor of 3 smaller than $\lambda_a^{(inst)}=-m_a^2/f_a^2$, the one extracted from the single cosine potential $V^{inst}(a)=-m_a^2 f_a^2 \cos(a/f_a)$. The six-axion couplings differ in sign as well. 

The VEV for the neutral pion, $\langle \pi^0 \rangle=\phi_a f_\pi$ can be shifted away by a non-singlet chiral rotation. Its presence is due to the $\pi^0$-$a$ mass mixing induced by isospin breaking effects in eq.~(\ref{eq:chpt-mass}), but can be avoided by a different choice for $Q_a$, which is indeed fixed up to a  non-singlet chiral rotation. As noticed in \cite{Georgi:1986df}, expanding eq.~(\ref{eq:chpt-mass}) to quadratic order in the fields we find the term
\begin{equation}
{\cal L}_{p^2} \supset 2 B_0 \frac{f_\pi}{4 f_a} a \langle \Pi \{Q_a,M_q\}\rangle ,
\end{equation}
which is responsible for the mixing. It is then enough to choose 
\begin{equation} \label{eq:Qa}
Q_a=\frac{M_q^{-1}}{\langle M_q^{-1} \rangle}\,,
\end{equation}
to avoid the tree-level mixing between the axion and pions and the VEV for the latter.
Such a choice only works at tree level, the mixing reappears at the loop level, 
but this contribution is small and can be treated as a perturbation.

The non-trivial potential (\ref{eq:pota}) allows for domain wall solutions. 
These have width ${\cal O}(m_a^{-1})$ and tension given by
\begin{equation}
\sigma = 8 m_a f_a^2\ {\cal E}\hspace{-3pt} \left [ \frac{4 m_u m_d}{(m_u+m_d)^2} \right]\,,
\qquad {\cal E}[q]\equiv \int_0^1 \frac{dy}{\sqrt{2(1-y)(1-q y)}}\,.
\end{equation}
The function ${\cal E}[q]$ can be written in terms of elliptic functions 
but the integral form is more compact.
Note that changing the quark masses over the whole possible range, $q\in [0,1]$, only varies
${\cal E}[q]$  between ${\cal E}[0]=1$ (cosine-like potential limit) 
and ${\cal E}[1]=4-2\sqrt2\simeq 1.17$ (for degenerate quarks). 
For physical quark masses ${\cal E}[q_{phys}]\simeq 1.12$, only 12\% off 
the cosine potential prediction, and $\sigma\simeq 9 m_a f_a^2$.

In a non vanishing axion field background, such as inside the domain wall or to a much lesser extent 
in the axion dark matter halo, QCD properties are different than in the vacuum. 
This can easily be seen expanding eq.~(\ref{eq:potap}) at the quadratic order in the pion field. 
For $\langle a \rangle=\theta f_a\neq 0$
the pion mass becomes
\begin{equation} \label{eq:mpitheta}
m_\pi^2(\theta)= m_\pi^2 \sqrt{1-\frac{4 m_u m_d}{(m_u+m_d)^2}\sin^2 \left(\frac{\theta}{2}\right) }\,,
\end{equation}
and for $\theta=\pi$ the pion mass is reduced by a factor $\sqrt{(m_d+m_u)/(m_d-m_u)}\simeq \sqrt3$.
Even more drastic effects are expected to occur in nuclear physics (see e.g.~\cite{Ubaldi:2008nf}).

The axion coupling to photons can also be reliably extracted from the chiral Lagrangian. Indeed at leading order it can simply be read out of eqs.~(\ref{eq:Laga2}), (\ref{eq:Laga2def}) and
(\ref{eq:Qa})\footnote{The result can also be obtained using a different choice of $Q_a$, but in
this case the non-vanishing $a$-$\pi^0$ mixing would require the inclusion of an extra contribution from the $\pi^0 \gamma \gamma$ coupling.}:
\begin{equation} \label{eq:gaggLO}
g_{a\gamma\gamma}= 
\frac{\alpha_{em}}{2\pi f_a} \left[\frac{E}{N} - \frac23\,\frac{4 m_d+m_u}{m_d+m_u}\right]\,,
\end{equation}
where the first term is the model dependent contribution proportional to the EM anomaly of the PQ symmetry, while the second is the model independent one coming from the minimal coupling to QCD at the non-perturbative level.

The other axion couplings to matter are either more model dependent (as the derivative couplings) or theoretically more challenging to study (as the coupling to EDM operators), or both. In section~\ref{sec:othercouplings}, we present a new strategy to extract the axion couplings to nucleons using experimental data and lattice QCD simulations. Unlike previous studies our
analysis is based only on first principle QCD computations. 
 While the precision is not as good as for the coupling to photons, 
 the uncertainties are already below $10\%$ and may improve as more lattice
simulations are performed.

Results with the 3-flavor chiral Lagrangian are often found in the literature. 
In the 2-flavor Lagrangian the extra contributions from the strange quark 
are contained inside the low-energy couplings. Within the 2-flavor effective 
theory the difference between using 2 or 3 flavor formulae,  is a higher order effect. 
Indeed the difference is ${\cal O}(m_u/m_s)$ which corresponds to the expansion 
parameter of the 2-flavor Lagrangian. As we will see in the next section these 
effects can only be consistently considered after including the full NLO correction.

At this point the natural question is, how good are the estimates obtained so far 
using leading order chiral Lagrangians? In the 3-flavor chiral Lagrangian NLO 
corrections are typically around 20-30\%.
The 2-flavor theory enjoys a much better perturbative expansion given 
the larger hierarchy between pions and the other mass thresholds. 
To get a quantitative answer the only option is to perform a complete NLO computation. 
Given the better behaviour of the 2-flavor expansion we perform
all our computation with the strange quark integrated out. 
The price we pay is the reduced number of physical observables 
that can be used to extract the higher order couplings. 
When needed we will use the 3-flavor theory to extract the values 
of the 2-flavor ones. This will produce intrinsic uncertainties ${\cal O}(30\%)$
in the extraction of the 2-flavor couplings. Such uncertainties however will
only have a small impact on the final result whose dependence on the higher order
2-flavor couplings is suppressed by the light quark masses.

\subsection{The mass}
The first quantity we compute is the axion mass. As mentioned before at leading order 
in $1/f_a$ the axion can be treated as an external source. Its mass is thus defined as
\begin{equation}
m_a^2= \frac{\delta^2 }{\delta a^2} \log {\cal Z}\bigl({\textstyle\frac{a}{f_a}}\bigr)\Bigr|_{a=0}= 
	\frac{1}{f_a^2}\frac{d^2 }{d \theta^2} \log {\cal Z}(\theta)\Bigr|_{\theta=0}
		= \frac{\chi_{top}}{f_a^2}\,,
\end{equation}
where  ${\cal Z}(\theta)$ is the QCD generating functional in the presence 
of a theta term  and $\chi_{top}$ is the topological susceptibility.

A partial computation of the axion mass at one loop was first attempted in \cite{Spalinski:1988az}.
More recently the full NLO corrections to $\chi_{top}$ has been computed in \cite{Mao:2009sy}.
We recomputed this quantity independently and present the result for the axion mass directly in terms of observable renormalized quantities\footnote{The results in \cite{Mao:2009sy} are instead
presented in terms of the unphysical masses and couplings in the chiral limit. Retaining the full
explicit dependence on the quark masses those formula are more suitable for lattice simulations.}.

The computation is very simple but the result has interesting properties:
\begin{equation} \label{eq:mares}
m_a^2=\frac{m_u m_d}{(m_u+m_d)^2}\frac{m_\pi^2 f_\pi^2}{f_a^2}\left[ 
1+2\frac{m_\pi^2}{f_\pi^2} \left(h_1^r-h_3^r-l_4^r
+ \frac{m_u^2-6m_u m_d + m_d^2}{(m_u+m_d)^2} l_7^r\right) \right]\,,
\end{equation} 
where $h_1^r$, $h_3^r$, $l_4^r$ and $l_7^r$ are the renormalized NLO couplings of \cite{Gasser:1983yg} and $m_\pi$ and $f_\pi$ are the physical (neutral) pion mass and decay constant (which include NLO corrections).
There is no contribution from loop diagrams at this order (this is true only after having 
reabsorbed the one loop corrections of the tree-level factor $m_\pi^2 f_\pi^2$). 
In particular $l_7^r$ and the combinations $h_1^r-h_3^r-l_4^r$ are separately scale invariant.
Similar properties are also present in the 3-flavor computation, in particular there are no ${\cal O}(m_s)$ corrections 
(after renormalization of the tree-level result), as noticed already in \cite{Spalinski:1988az}.

To get a numerical estimate of the axion mass and the size of the corrections
we need the values of the NLO couplings. In principle $l_7^r$ could be extracted from
the QCD contribution to the $\pi^+$-$\pi^0$ mass splitting. While lattice simulations have
 started to become sensitive to EM and isospin breaking effects, at the moment there are 
no reliable estimates of this quantity from first principle QCD. Even less is known
about $h_1^r-h_3^r$, which does not enter other measured observables. The only hope would
be to use lattice QCD computation to extract such coupling by studying the quark mass
dependence of observables such as the topological susceptibility. Since these studies
are not yet available we employ a small trick: we use the relations in \cite{Gasser:1984gg} 
between the 2- and 3-flavor couplings to circumvent the problem. In particular we have
\begin{align} \label{eq:l7}
l_7^r & = \frac{m_u+m_d}{m_s}\frac{f_\pi^2}{8 m_\pi^2}-36 L_7-12 L_8^r +\frac{\log(m_\eta^2/\mu^2)+1}{64\pi^2}+\frac{3\log(m_K^2/\mu^2)}{128\pi^2} = 7(4) \cdot 10^{-3}\,, \nonumber \\
h_1^r-h_3^r-l_4^r & = -8L_8^r+\frac{\log(m_\eta^2/\mu^2)}{96\pi^2}+\frac{\log(m_K^2/\mu^2)+1}{64\pi^2} = (4.8\pm 1.4) \cdot 10^{-3}\,.
\end{align}
The first term in $l_7^r$ is due to the tree-level contribution to the $\pi^+$-$\pi^0$ mass splitting 
due to the $\pi^0$-$\eta$ mixing from isospin breaking effects. 
The rest of the contribution, formally NLO, includes the effect of the $\eta$-$\eta'$ mixing 
and numerically is as important as the tree-level piece \cite{Gasser:1984gg}.
We thus only need the values of the 3-flavor couplings $L_7$ and $L_8^r$, which can be extracted from 
chiral fits \cite{Bijnens:2014lea} and lattice QCD \cite{Aoki:2013ldr}, 
we refer to appendix~\ref{sec:app1} for more details on the
values used. An important point is that by using 3-flavor couplings the precision of the estimates
of the 2-flavor ones will be limited to the convergence of the 3-flavor Lagrangian. However, 
given the small size of such corrections even an ${\cal O}(1)$ uncertainty will still translate into
a small overall error.

The final numerical ingredient needed is the actual up 
and down quark masses, in particular their ratio. 
Since this quantity already appears  in the tree level formula of the axion mass
we need a precise estimate for it, however,
because of the Kaplan-Manohar (KM) ambiguity \cite{Kaplan:1986ru},
it cannot be extracted within the meson Lagrangian. 
Fortunately recent lattice QCD simulations have dramatically improved 
our knowledge of this quantity. Considering the latest results we take
\begin{equation} \label{eq:defz}
z\equiv \frac{m_u^{\overline {\rm MS}}(2~{\rm GeV})}{m_d^{\overline {\rm MS}}(2~{\rm GeV})}=0.48(3)\,,
\end{equation}
where we have conservatively taken a larger error than the one coming from simply averaging
the results in \cite{deDivitiis:2013xla,Basak:2015lla,Horsley:2015eaa} 
(see the appendix~\ref{sec:app1} for more details). 
Note that $z$ is scale independent up to $\alpha_{em}$ and Yukawa suppressed
corrections. Note also that since lattice QCD simulations allow us to relate physical observables
directly to the high-energy $\overline {\rm MS}$ Yukawa couplings,  
in principle\footnote{Modulo well-known effects present when chiral non-preserving
fermions are used.}, they do not suffer from the KM ambiguity, which is a feature of chiral Lagrangians.
It is reasonable to expect that the precision on the ratio $z$ will increase further in the near future. 

Combining everything together we get the following numerical estimate for the axion mass
\begin{align}
m_a & = 5.70(6)(4)~\mu{\rm eV}\,\left(\frac{10^{12}{\rm GeV}}{f_a}\right)
	=5.70(7)~\mu{\rm eV}\,\left(\frac{10^{12}{\rm GeV}}{f_a}\right) \,,
\end{align}
where the first error comes from the up-down quark mass ratio uncertainties (\ref{eq:defz}) while the second
comes from the uncertainties in the low energy constants (\ref{eq:l7}). The total error of $\sim$1\% is much smaller than
the relative errors in the quark mass ratio ($\sim$6\%) and in the NLO couplings ($\sim$30$\div$60\%) because of the weaker dependence of the axion mass on these quantities
\begin{align}
m_a=\left[5.70+0.06\, \frac{z-0.48}{0.03}
-0.04\, \frac{10^3 l_7^r-7}{4} 
+0.017\, \frac{10^3(h_1^r-h_3^r-l_4^r)-4.8}{1.4} \right]\mu{\rm eV}\,\frac{10^{12}\,{\rm GeV}}{f_a} \,.
\end{align}
Note that the full NLO correction is numerically smaller than the quark mass error and 
its uncertainty is dominated by $l_7^r$. The error on the latter is particularly
large because of a partial cancellation between $L_7^r$ and $L_8^r$ in eq.~(\ref{eq:l7}). 
The numerical irrelevance of the other NLO couplings leaves a lot of room for improvement
should $l_7^r$ be extracted directly from Lattice QCD.  

The value of the pion decay constant  we used ($f_\pi=92.21(14)$~MeV) 
\cite{Agashe:2014kda} is extracted 
from $\pi^+$~decays and includes the leading
QED corrections, other ${\cal O}(\alpha_{em})$  corrections to $m_a$ are expected 
to be sub-percent. Further reduction of the error on the axion mass may require a dedicated study 
of this source of uncertainty as well.

As a by-product we also provide a comparably high precision estimate of 
the topological susceptibility itself
\begin{equation}
\chi_{top}^{1/4}=\sqrt{m_a f_a}= 75.5(5)\,{\rm MeV}
\,,
\end{equation}
against which lattice simulations can be calibrated.

\subsection{The potential: self-coupling and domain-wall tension}
Analogously to the mass, the full axion potential can be straightforwardly computed at NLO. 
There are three contributions: the pure Coleman-Weinberg
1-loop potential from pion loops, the tree-level contribution from the NLO Lagrangian, and
the corrections from the renormalization of the tree-level result, when rewritten in terms of
physical quantities ($m_\pi$ and $f_\pi$). The full result is
\begin{align} \label{eq:VaNLO}
V(a)^{\rm NLO} = & -m_{\pi}^2\bigl({\textstyle\frac{a}{f_a}}\bigr) f_\pi^2 \left \{
1-2\frac{m_\pi^2}{f_\pi^2} \left[ l_3^r+l_4^r-\frac{(m_d-m_u)^2}{(m_d+m_u)^2} l_7^r-\frac{3}{64\pi^2}\log \left(\frac{m_\pi^2}{\mu^2}\right)\right] \nonumber \right.  \\
+ & \left.     \frac{m_{\pi}^2\bigl({\textstyle\frac{a}{f_a}}\bigr)}{f_\pi^2} \left [
h_1^r-h_3^r+l_3^r+\frac{4 m_u^2 m_d^2}{(m_u+m_d)^4}\frac{m_\pi^8 \sin^2\bigl({\textstyle\frac{a}{f_a}}\bigr)}{m_{\pi}^8\bigl({\textstyle\frac{a}{f_a}}\bigr)}l_7^r-\frac{3}{64\pi^2}\left(\log \left(\frac{m_\pi^2\bigl({\textstyle\frac{a}{f_a}}\bigr)}{\mu^2}\right)-\frac12 \right)
\right] \right\} 
\end{align}
where $m_\pi^2(\theta)$  is the function defined in eq.~(\ref{eq:mpitheta}), and all quantities have been rewritten
in terms of the physical NLO quantities\footnote{See also \cite{Guo:2015oxa} for a related result computed in terms of the LO quantities.}. In particular the first line comes from the NLO corrections of the 
tree-level potential while the second line is the pure NLO correction to the effective potential.

The dependence on the axion is highly non-trivial, however the NLO corrections account for only up to few
percent change in the shape of the potential (for example the difference in vacuum energy between the minimum
and the maximum of the potential changes by 3.5\% when NLO corrections are included).
The numerical values for the additional low-energy constants $l_{3,4}^r$ are reported in appendix~\ref{sec:app1}. We thus know the full QCD axion potential at the percent level!

It is now easy to extract the self-coupling of the axion at NLO by expanding the effective potential
(\ref{eq:VaNLO}) around the origin
\begin{equation}
V(a)= V_0+ \frac12 m_a^2 a^2 + \frac{\lambda_a}{4!} a^4 +\dots
\end{equation}
We find
\begin{align}
\lambda_a = & - \frac{m_a^2}{f_a^2} \left \{ \frac{m_u^2-m_u m_d+m_d^2}{(m_u+m_d)^2} \right.
 \\ &  \left. +6\frac{m_\pi^2}{f_\pi^2}
\frac{m_u m_d}{(m_u+m_d)^2} \left[ h_1^r-h_3^r-l_4^r+\frac{4 \bar{l}_4-\bar l_3-3}{64\pi^2}
-4 \frac{m_u^2-m_u m_d+m_d^2}{(m_u+m_d)^2} l_7^r
 \right] \right \}\,,
\end{align}
where $m_a$ is the physical one-loop corrected axion mass of eq.~(\ref{eq:mares}). Numerically we have
\begin{equation}
\lambda_a=-0.346(22)\cdot \frac{m_a^2}{f_a^2}\,,
\end{equation}
the error on this quantity amounts to roughly 6\% and is dominated by the uncertainty on $l_7^r$.

Finally the NLO result for the domain wall tensions can be simply extracted from the definition
\begin{equation}
\sigma=2  f_a \int_0^\pi  d \theta\,  \sqrt{2[V(\theta)-V(0)]}\,,
\end{equation}
using the NLO expression (\ref{eq:VaNLO}) for the axion potential.
The numerical result is
\begin{equation}
\sigma=8.97(5)\,m_a f_a^2 \,,
\end{equation}
the error is sub percent and it receives comparable contributions from the errors 
on $l_7^r$ and the quark masses.

As a by-product we also provide a precision estimate of the topological quartic moment of the
topological charge $Q_{top}$
\begin{equation}
b_2\equiv -\frac{\langle Q_{top}^4 \rangle-3\langle Q_{top}^2 \rangle^2 }{12\langle Q_{top}^2 \rangle}=
\frac{f_a^2 V''''(0)}{12 V''(0)}=\frac{\lambda_a f_a^2 }{12 m_a^2}=-0.029(2)\,,
\end{equation}
to be compared to the cosine-like potential $b_2^{inst}=-1/12\simeq -0.083$.

\subsection{Coupling to photons}
Similarly to the axion potential, the coupling to photons (\ref{eq:gaggLO}) also gets
QCD corrections at NLO, which are completely model independent. 
Indeed derivative couplings only produce $m_a$ suppressed corrections which
are negligible, thus the only model dependence lies in the anomaly
coefficient $E/N$.

For physical quark masses the QCD contribution (the second term in eq.~(\ref{eq:gaggLO}))
is accidentally close to $-2$. This implies that models with $E/N=2$ can 
have anomalously small coupling to photons, relaxing astrophysical bounds. 
The degree of this cancellation is very sensitive to the uncertainties from the quark mass 
and the higher order corrections, which we compute here for the first time.

At NLO new couplings appear from higher-dimensional operators correcting the WZW Lagrangian. Using the basis of \cite{Bijnens:2001bb}, the result reads
\begin{align} \label{eq:gaggNLO}
g_{a\gamma \gamma}=\frac{\alpha_{em}}{2\pi f_a} \left \{ 
\frac{E}{N}-\frac{2}{3}\frac{4m_d+m_u}{m_d+m_u}+\frac{m_\pi^2}{f_\pi^2}\frac{8 m_u m_d}{(m_u+m_d)^2} \left[
\frac89 \left( 5 \tilde c^W_3+\tilde c^W_7+2\tilde c^W_8 \right)
-\frac{m_d-m_u}{m_d+m_u}l_7^r
\right]
\right\}\,.
\end{align}
The NLO corrections in the square brackets come from tree-level diagrams with insertions of NLO WZW operators  (the terms proportional to the $\tilde c^W_i$ couplings\footnote{For simplicity we have rescaled the original couplings $c_i^W$ of \cite{Bijnens:2001bb} into  $\tilde c_i^W\equiv c_i^W(4\pi f_\pi)^2$.}) and from $a$-$\pi^0$ mixing diagrams
(the term proportional to $l_7^r$). One loop diagrams exactly cancel similarly to
what happens for $\pi\to\gamma\gamma$ and $\eta\to\gamma\gamma$ \cite{Donoghue:1986wv}.
Notice that the $l_7^r$ term includes the $m_u/m_s$ contributions 
which one obtains from the 3-flavor tree-level computation. 

Unlike the NLO couplings entering the axion mass and potential little is known about
the couplings $\tilde c_i^W$, so we describe  the way to extract them here. 

The first obvious observable we can use is the $\pi^0\to\gamma\gamma$ width. 
Calling $\delta_i$ the relative correction at NLO to the amplitude for the $i$ process, i.e.
\begin{equation}
\Gamma_i^{\rm NLO} \equiv \Gamma_i^{\rm tree} (1+\delta_i)^2\,,
\end{equation} 
the expressions for $\Gamma_{\pi\gamma\gamma}^{\rm tree}$ and $\delta_{\pi\gamma\gamma}$ read
\begin{align} \label{eq:pigg2}
\Gamma_{\pi\gamma\gamma}^{\rm tree}&=\frac{\alpha_{em}^2}{(4\pi)^3}\frac{m_\pi^3}{f_\pi^2}\,,\quad\ \delta_{\pi\gamma\gamma}  = \frac{16}{9} \frac{m_\pi^2}{f_\pi^2}\left[ 
\frac{m_d-m_u}{m_d+m_u} \left( 5 \tilde c^W_3+\tilde c^W_7+2\tilde c^W_8 \right)
- 3 \left( \tilde c^W_3 + \tilde c^W_7+ \frac{\tilde c^W_{11}}{4} \right)
\right] . 
\end{align}
Once again the loop corrections are reabsorbed by the renormalization of the tree-level parameters
and the only contributions come from the NLO WZW terms. While the isospin breaking correction involves
exactly the same combination of couplings entering the axion width, the isospin preserving one does
not. This means that we cannot extract
the required NLO couplings from the pion width alone. However in the absence of large cancellations
between the isospin breaking and the isospin preserving contributions we can use the experimental value for the pion decay rate to estimate the order of magnitude of the corresponding corrections to the axion case.
Given the  small difference between the experimental and the tree-level prediction for $\Gamma_{\pi\to\gamma\gamma}$ the NLO axion correction is expected of order few percent.

To obtain numerical values for the unknown couplings we can try to use the 3-flavor theory, in analogy with the axion mass computation. In fact at NLO in the 3-flavor theory the decay rates $\pi\to\gamma\gamma$ and $\eta\to\gamma\gamma$ only depend on two low-energy couplings that can thus be determined. 
Matching these couplings to the 2-flavor theory ones we are able to extract the required combination
entering in the axion coupling. Because the ${\tilde c}_i^W$ couplings enter eq.~(\ref{eq:gaggNLO}) only at NLO in the light quark mass expansion we only need to determine them at LO in the $m_{u,d}$ expansion.

The $\eta\to\gamma\gamma$ decay rate at NLO is
\begin{align}
\Gamma_{\eta\to\gamma\gamma}^{\rm tree}&= 
\frac{\alpha^2_{em}}{3(4\pi)^3}\frac{m_\eta^3}{f_\eta^2}\,, \nonumber \\
\delta_{\eta\gamma\gamma}^{(3)} & =\frac{32}{9}\frac{m_\pi^2}{f_\pi^2}\left[
\frac{2m_s-4m_u-m_d}{m_u+m_d}\tilde C^W_7+6\,\frac{2m_s-m_u-m_d}{m_u+m_d}\tilde C^W_8
\right]  \simeq  \frac{64}{9}\frac{m_K^2}{f_\pi^2} \left( \tilde C^W_7+6\, \tilde C^W_8 \right)\,,
\end{align}
where in the last step we consistently neglected higher order corrections ${\cal O}(m_{u,d}/m_s)$.
The 3-flavor couplings $\tilde C_i^W\equiv (4\pi f_\pi)^2 C_i^W$ are defined in \cite{Bijnens:2001bb}.
The expression for the correction to the $\pi\to\gamma\gamma$ amplitude with 3 flavors also receives important corrections from the $\pi$-$\eta$ mixing $\epsilon_2$,
\begin{align} \label{eq:pigg3}
\delta_{\pi\gamma\gamma}^{(3)} & =
\frac{32}{9}\frac{m_\pi^2}{f_\pi^2}\left[
\frac{m_d-4m_u}{m_u+m_d}\tilde C^W_7+6\,\frac{m_d-m_u}{m_u+m_d}\tilde C^W_8
\right]+\frac{f_\pi}{f_\eta}\frac{\epsilon_2}{\sqrt3}(1+\delta_{\eta\gamma\gamma})\,,
\end{align}
where the $\pi$-$\eta$ mixing derived in \cite{Gasser:1984gg} can be conveniently rewritten as
\begin{equation}
\frac{\epsilon_2}{\sqrt3}\simeq\frac{m_d-m_u}{6m_s}\left[ 
1+\frac{4m_K^2}{f_\pi^2}\left (l_7^r-\frac{1}{64\pi^2}\right) \right]\,,
\end{equation}
at leading order in $m_{u,d}$.
In both decay rates the loop corrections are reabsorbed in the renormalization of the tree-level amplitude\footnote{NLO corrections to $\pi$ and $\eta$ decay rates to photons
including isospin breaking effects were also computed  in \cite{Ananthanarayan:2002kj}. For the $\eta\to\gamma\gamma$ rate we disagree in the expression of the terms ${\cal O}(m_{u,d}/m_s)$, which are however subleading. For the  $\pi\to\gamma\gamma$ rate we also included the mixed term coming from the product
of the NLO corrections to $\epsilon_2$ and to $\Gamma_{\eta\gamma\gamma}$. 
Formally this term is NNLO but given that the NLO corrections to 
both $\epsilon_2$ and $\Gamma_{\eta\gamma\gamma}$ are
of the same size as the corresponding LO contributions
such terms cannot be neglected.
}.

\begin{figure}[t!]
\centering
\includegraphics[scale=0.4]{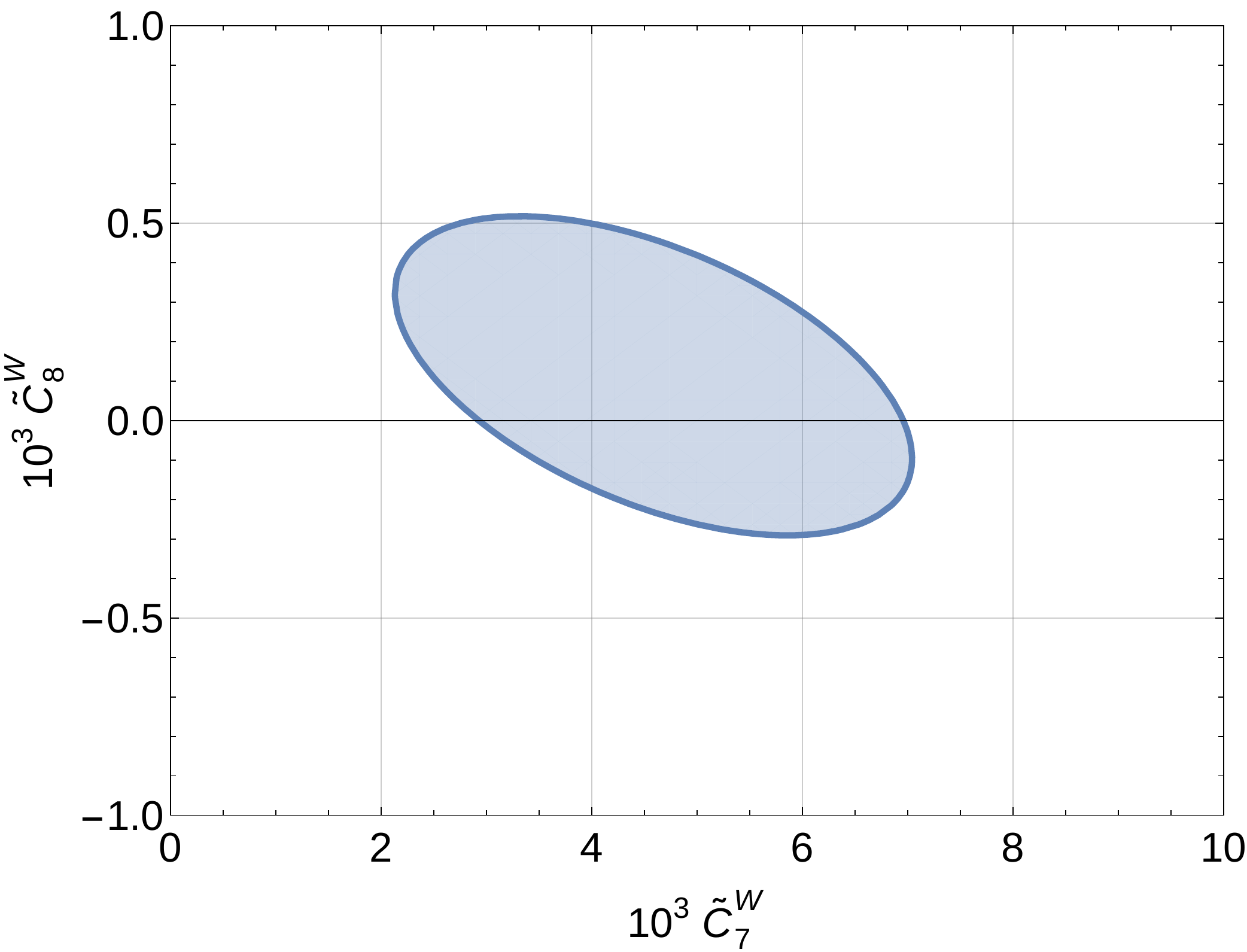}
\caption{Result of the fit of the 3-flavor couplings $\tilde C^W_{7,8}$ 
from the decay width of $\pi\to\gamma\gamma$ and $\eta\to\gamma\gamma$, which
include the experimental uncertainties and a 30\% systematic uncertainty from
higher order corrections. \label{fig:c7c8}}
\end{figure}

By comparing the light quark mass dependence in eqs.~(\ref{eq:pigg2}) and (\ref{eq:pigg3}) we can match the 2 and 3 flavor couplings as follows
\begin{align}
\tilde c_3^W+\tilde c_7^W+\frac{\tilde c_{11}^W}{4} & = \tilde C_7^W \,, \nonumber \\
5\tilde c_3^W+\tilde c_7^W+2\tilde c_{8}^W & =  5  \tilde C_7^W +12  \tilde C_8^W
+\frac{3}{32} \frac{f_\pi^2}{m_K^2} \left [ 1+4\frac{m_K^2}{f_\pi f_\eta}\left( l_7^r-\frac{1}{64\pi^2} \right) \right ]\left( 1+\delta_{\eta\gamma\gamma} \right) \,.
\end{align}
Notice that the second combination of couplings is exactly the one needed
for the axion-photon coupling.
By using the experimental results for the decay rates (reported in appendix~\ref{sec:app1}),
we can extract $C^W_{7,8}$. The result is shown in fig.~\ref{fig:c7c8},
the precision is low for two reasons: 1)  $\tilde C^W_{7,8}$
are 3 flavor couplings so they suffer from an intrinsic ${\cal O}$(30\%) uncertainty
from higher order corrections\footnote{We implement these uncertainties by adding 
a 30\% error on the experimental input values of $\delta_{\pi\gamma\gamma}$ and $\delta_{\eta\gamma\gamma}$.}, 2) for $\pi\to\gamma\gamma$ the experimental
uncertainty is not smaller than the NLO corrections we want to fit.

For the combination 
$5\tilde c_3^W+\tilde c_7^W+2\tilde c_{8}^W$ we are interested in, the final result reads
\begin{align}
5 \tilde c^W_3+\tilde c^W_7+2\tilde c^W_8  & =
\frac{3 f_\pi^2}{64 m_K^2}\frac{m_u+m_d}{m_u}
\left\{\left[1+4\frac{m_K^2}{f_\pi^2}\left(l_7^r-\frac{1}{64\pi^2}\right) \right]\frac{f_\pi}{f_\eta}(1+\delta_{\eta\gamma\gamma})+3\delta_{\eta\gamma\gamma}-6\frac{m_K^2}{m_\pi^2} \delta_{\pi\gamma\gamma}
 \right\} \nonumber \\
& = 0.033(6) \,.
\end{align}  
When combined with eq.~(\ref{eq:gaggNLO}) we finally get
\begin{align} \label{eq:gaggnum}
g_{a\gamma \gamma}=\frac{\alpha_{em}}{2\pi f_a} \left[ 
\frac{E}{N} -1.92(4)\right] = \left[ 0.203(3)\frac{E}{N}-0.39(1) \right] \frac{m_a}{\rm GeV^2} \,.
\end{align}
Note that despite the rather large uncertainties of the NLO couplings we are able to extract 
the model independent contribution to $a\to\gamma\gamma$ at the percent level. This is due
to the fact that, analogously to the computation of the axion mass, the NLO corrections are suppressed by
the light quark mass values. Modulo experimental uncertainties eq.~(\ref{eq:gaggnum}) would allow the parameter $E/N$ 
to be extracted  from a measurement of $g_{a\gamma\gamma}$ at the percent level.
\begin{figure}[t!]
\centering
\includegraphics[scale=0.4]{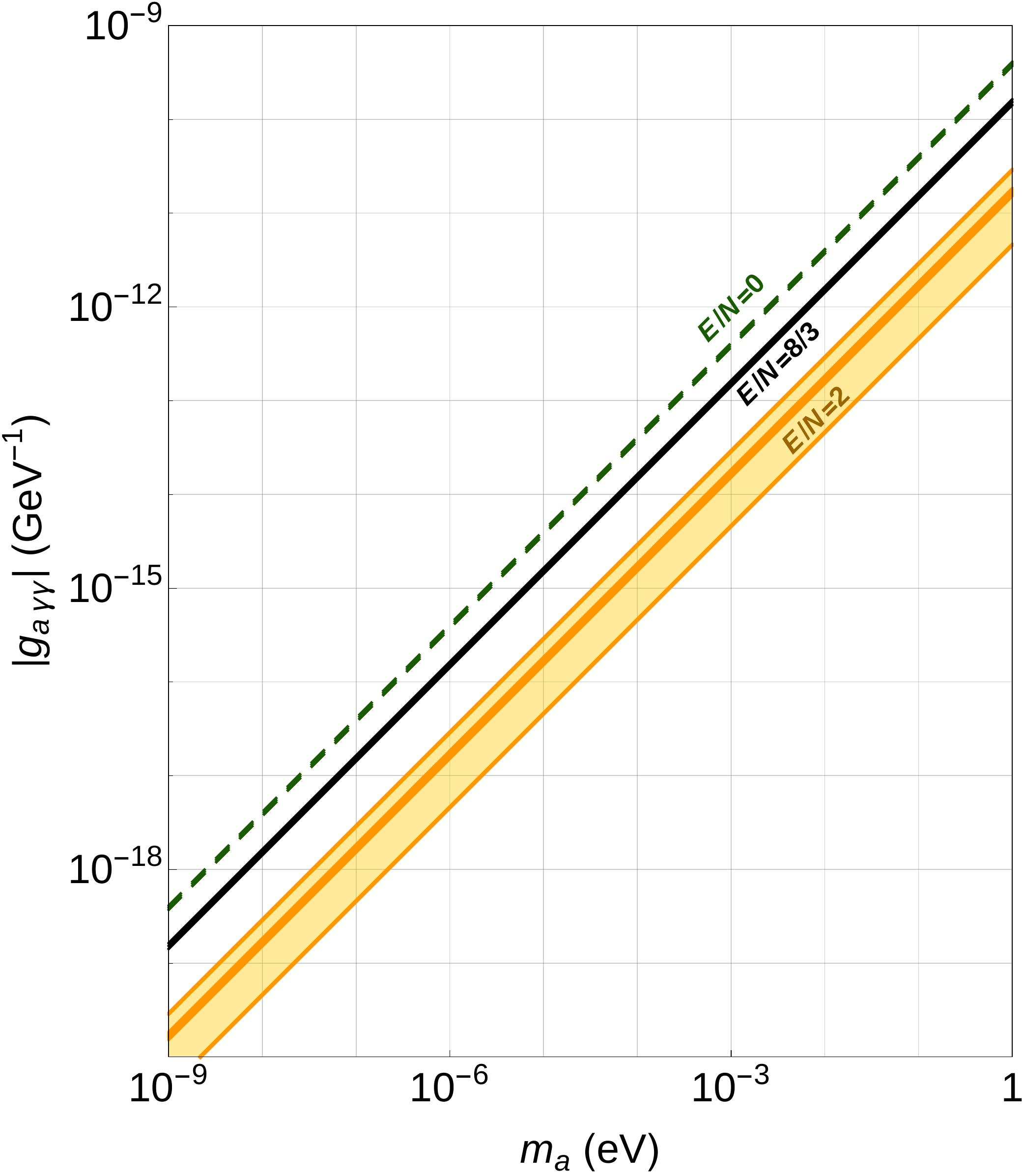}
\caption{The relation between the axion mass and its coupling to photons for the three reference models with $E/N=0$, $8/3$ and $2$. Notice the larger relative uncertainty in the latter model due to the cancellation between the UV and IR contributions to the anomaly (the band corresponds to $2\sigma$ errors.). Values below the lower band require a higher degree of cancellation. \label{fig:gavsma}}
\end{figure}

For the three reference models with respectively $E/N=0$ (such as hadronic or KSVZ-like models~\cite{Kim:1979if,Shifman:1979if} with electrically neutral heavy fermions), $E/N=8/3$ (as in DFSZ models~\cite{Zhitnitsky:1980tq,Dine:1981rt} or KSVZ models with heavy fermions in complete SU(5) representations) and $E/N=2$ (as in some KSVZ ``unificaxion'' models \cite{Giudice:2012zp}) the coupling reads
\begin{equation} \label{eq:gsols}
g_{a\gamma \gamma}=\left \{ \begin{array}{ll} -2.227(44) \cdot 10^{-3}/f_a & E/N=0 \\ 
\phantom{-}0.870(44)  \cdot 10^{-3}/f_a & E/N=8/3 \\ 
\phantom{-}0.095(44)  \cdot 10^{-3}/f_a & E/N=2
\end{array} \right . \,.
\end{equation}
Even after the inclusion of NLO corrections the coupling to photons in $E/N=2$ models is still suppressed.
The current uncertainties are not yet small enough to completely rule out a higher degree of cancellation, 
but a suppression bigger than ${\cal O}$(20) with respect to $E/N=0$ models
is highly disfavored. Therefore the result for $g_{a\gamma\gamma}^{E/N=2}$ of eq.~(\ref{eq:gsols}) can now be taken
 as a lower bound to the axion coupling to photons, below which tuning is required. 
 The result is shown in fig.~\ref{fig:gavsma}.

\subsection{Coupling to matter} \label{sec:othercouplings}
Axion couplings to matter are more model dependent as they depend on all the UV couplings defining the effective axial current (the constants $c_q^0$ in the last term of eq.~(\ref{eq:Laga})).
In particular, there is a model independent contribution coming from the axion coupling to gluons 
(and to a lesser extent to the other gauge bosons) and a model dependent part contained
in the fermionic axial couplings.

The couplings to leptons can be read off directly from the UV Lagrangian up to the one loop effects coming from the coupling to the EW gauge bosons.
The couplings to hadrons are more delicate because they involve matching hadronic to elementary quark physics. Phenomenologically the most interesting ones are the axion couplings to nucleons, which could in principle be tested from long range force experiments, or from dark-matter direct-detection like experiments.

In principle we could attempt to follow a similar procedure to the one used in the previous section,
namely to employ chiral Lagrangians with baryons and use known experimental data to extract the necessary 
low energy couplings. 
Unfortunately effective Lagrangians involving baryons are on much less solid ground---there are no
parametrically large energy gaps in the hadronic spectrum to justify the use of low energy expansions.

A much safer thing to do is to use an effective theory valid at energies much lower
than the QCD mass gaps $\Delta\sim{\cal O}$(100~MeV). In this regime nucleons are non-relativistic,
their number is conserved and they can be treated as external fermionic currents.
For exchanged momenta $q$ parametrically 
smaller than $\Delta$, heavier modes are not excited and the effective field theory is under control. 
The axion, as well as the electro-weak gauge bosons, enters as classical sources in the effective Lagrangian, which would otherwise be a free non-relativistic Lagrangian at leading order. 
At energies much smaller than the QCD mass gap the only active flavor symmetry we can use is isospin, which is explicitly broken only by the small quark masses (and QED effects). The leading order effective Lagrangian for the 1-nucleon sector reads 
\begin{equation} \label{eq:lagN}
 {\cal L}_{N}=\bar N v^\mu D_\mu N+2 g_A\,   A^i_\mu\,  \bar N S^\mu  \sigma^i  N
 + 2 g_0^q\, {\hat A}^q_\mu \,\bar N S^\mu N   + \sigma \langle M_a \rangle \bar N N +b \bar N M_a N+\dots 
 \end{equation} 
where $N=(p,n)$ is the isospin doublet nucleon field,  
$v^\mu$ is the four-velocity of the non-relativistic nucleons,
 $D_\mu=\partial_\mu -V_\mu$,  $V_\mu$ is the vector external current,
$\sigma^i$ are the Pauli matrices, the index $q=(\frac{u+d}2,s,c,b,t)$ runs
over isoscalar quark combinations,
$2\bar N S^\mu N=\bar N  \gamma^\mu \gamma_5 N$ is the nucleon axial current,
$M_a=\cos(Q_a a/f_a){\rm diag}(m_u,m_d)$,
and ${A}^{i}_\mu$ and $\hat A^{q}_\mu$
are the axial isovector and isoscalar external currents respectively.
Neglecting SM gauge bosons, the external currents only depend on the axion field as
follows
\begin{align}
\hat A_\mu^{q}  =c_q \frac{\partial_\mu a}{2f_a}\,, \qquad 
A^3_\mu  =c_{(u-d)/2}\ \frac{\partial_\mu a}{2f_a}\,, \qquad 
A_\mu^{1,2}=V_\mu=0\,,
\end{align}
where we used the short-hand notation $c_{(u\pm d)/2}\equiv\frac{c_u\pm c_d}{2}$.
The couplings $c_q=c_q(Q)$ computed at the scale $Q$ will in general differ
from the high scale ones because of the running of the anomalous axial current~\cite{Kodaira:1979pa}. 
In particular under RG evolution the couplings $c_q(Q)$ mix, so
that in general they will all be different from zero at low energy. 
We explain the details of this effect in appendix~\ref{app:Zrge}.

Note that the linear axion couplings to nucleons
are all contained in the derivative interactions through $A_\mu$ while there are no linear 
interactions\footnote{This is no longer true in the presence of extra CP violating operators such
as those coming from the CKM phase or new physics. The former are known to be very small,
while the latter are more model dependent and we will not discuss them in the current work.} coming from the non derivative terms contained in $M_a$.
In Eq.~(\ref{eq:lagN}) dots stand for higher order terms involving higher powers of the external sources $V_\mu$, $A_\mu$, 
and $M_a$. Among these the leading effects to the axion-nucleon coupling will come from
isospin breaking terms ${\cal O}(M_a A_\mu)$.\footnote{Axion 
couplings to EDM operators also appear at this order.} 
These corrections are small
${\cal O}(\frac{m_{d}-m_{u}}{\Delta})$, below the uncertainties associated to our
determination of the effective coupling $g_0^q$, which are extracted from lattice simulations
performed in the isospin limit.

Eq.~(\ref{eq:lagN}) should not be confused with the usual heavy baryon chiral Lagrangian~\cite{Jenkins:1990jv}  because here pions have 
been integrated out. The advantage of using this Lagrangian is clear: for axion physics 
the relevant scale is of order $m_a$, so higher order terms are negligibly small ${\cal O}(m_a/\Delta)$. The price to pay is that the couplings $g_A$ and $g_0^q$ can only be extracted from very low-energy experiments or lattice QCD simulations.  Fortunately the combination of the two will be enough for our purposes.

In fact, at the leading order in the isospin breaking expansion, $g_A$ and $g_0^q$  can simply be extracted  by matching single nucleon matrix elements computed with the QCD+axion Lagrangian (\ref{eq:Laga2}) and with the effective axion-nucleon theory (\ref{eq:lagN}). The result is simply:
\begin{align}
g_A  =\Delta u& -\Delta d\,,   \quad
g_0^q= ( \Delta u+\Delta d, \Delta s, \Delta c, \Delta b, \Delta t)\,, \quad 
 s^\mu \Delta q \equiv \langle p | \bar q \gamma^\mu \gamma_5 q | p \rangle\,,
\end{align}
where $|p \rangle$ is a proton state at rest, $s^\mu$ its spin and we used isospin symmetry
to relate proton and neutron matrix elements.
Note that the isoscalar matrix elements $\Delta q$ inside $g_0^q$ depend on the matching 
scale $Q$, such dependence is however canceled once the couplings $g_0^q(Q)$ are multiplied
by the corresponding UV couplings $c_q(Q)$ inside the isoscalar currents $\hat A^q_\mu$.
Non-singlet combinations such as $g_A$ are instead protected by non-anomalous
Ward identities\footnote{This is only true in renormalization schemes 
which preserve the Ward identities.}. 
For future convenience we set the matching scale $Q=2$~GeV.

We can therefore write the EFT Lagrangian (\ref{eq:lagN}) directly in terms of the UV couplings as
\begin{align} \label{eq:lagN2}
{\cal L}_N  =\bar N v^\mu D_\mu N
+\frac{\partial_\mu a}{f_a} \Bigl \{ &
\frac{c_u-c_d}2 (\Delta u-\Delta d)\bar N S^\mu \sigma^3  N  \nonumber \\
&  +\Bigl[ \frac{c_u+c_d}{2}(\Delta u+\Delta d) 
	+ \sum_{q=s,c,b,t} c_q \Delta q \Bigr] \bar N S^\mu   N
\Bigr \} \,.
\end{align}

We are thus left to determine the matrix elements $\Delta q$. The isovector combination 
can be obtained with high precision from $\beta$-decays~\cite{Agashe:2014kda}
\begin{equation} \label{eq:gAnum}
\Delta u - \Delta d=g_A=1.2723(23)\,,
\end{equation}
where the tiny neutron-proton mass splitting $m_n-m_p=1.3$~MeV guarantees that we are within
the regime of our effective theory. The error quoted is experimental and does not include possible
isospin breaking corrections.

Unfortunately we do not have other low energy experimental 
inputs to determine the remaining matrix elements. 
Until now such information has been extracted from
a combination of deep-inelastic-scattering data and semi-leptonic hyperon decays:
The former suffer from uncertainties coming from the integration over the low-$x$ kinematic 
region, which is known to give large contributions to the observable of interest; 
the latter are not really within the EFT regime, which does not allow a reliable estimate
of the accuracy. 

Fortunately lattice simulations have recently started producing direct 
reliable results for these matrix elements. From \cite{QCDSF:2011aa,Engelhardt:2012gd,
Abdel-Rehim:2013wlz,Bhattacharya:2015gma,Abdel-Rehim:2015owa,Abdel-Rehim:2015lha} 
(see also \cite{Bhattacharya:2013ehc,Yamanaka:2015lat}) we extract\footnote{Details in the way
the numbers in eq.~(\ref{eq:g0num}) are derived are given in appendix \ref{sec:app1}.} the
following inputs computed at $Q=2$~GeV in $\overline{\rm MS}$
\begin{align} \label{eq:g0num}
g_0^{ud}=\Delta u+\Delta d=0.521(53)\,,\qquad  \Delta s=-0.026(4)\,,\qquad \Delta c=\pm 0.004 \,.
\end{align}
Notice that the charm spin content is so small that its value has not been determined yet, only
an upper bound exists. Similarly we can neglect the analogous contributions from bottom and top quarks which are expected to be even smaller. As mentioned before, lattice simulations do not
include isospin breaking effects, these are however expected to be smaller than the current uncertainties. Combining eqs.~(\ref{eq:gAnum}) and (\ref{eq:g0num}) we thus get:
\begin{equation}
\Delta u=0.897(27)\,, \qquad \Delta d=-0.376(27)\,,\qquad \Delta s=-0.026(4)\,,
\end{equation}
computed at the scale $Q=2$~GeV.

We can now use these inputs in the EFT Lagrangian (\ref{eq:lagN2}) to extract the corresponding axion-nucleon couplings:
\begin{align} \label{eq:cpcn}
c_p & = -0.47(3)+0.88(3) c_u^0 -0.39(2) c_d^0-0.038(5)c_s^0 \nonumber \\
				&\hspace{2.15cm}-0.012(5) c_c^0 -0.009(2) c_b^0-0.0035(4) c_t^0\,, \nonumber  \\
c_n & = -0.02(3)+0.88(3) c_d^0 -0.39(2) c_u^0-0.038(5)c_s^0 \nonumber \\
				&\hspace{2.15cm}-0.012(5) c_c^0 -0.009(2) c_b^0-0.0035(4) c_t^0 \,,
\end{align}
which are defined in analogy to the couplings to quarks as
\begin{equation}
\frac{\partial_\mu a}{2 f_a}c_{N} \bar N \gamma^\mu \gamma_5 N \,, 
\end{equation}
and are scale invariant (as they are defined in the effective theory below the QCD mass gap). 
The errors in eq.~(\ref{eq:cpcn}) include the uncertainties from the lattice data
and those from higher order corrections in the perturbative RG evolution of the axial current
(the latter is only important for the coefficients of $c^0_{s,c,b,t}$). 
The couplings $c^0_q$ are those appearing in eq.~(\ref{eq:Laga}) computed at the high scale 
$f_a=10^{12}$~GeV. The effect of varying the matching scale to a different value of $f_a$
within the experimentally allowed range is smaller than the theoretical uncertainties.

A few considerations are in order. The theoretical errors quoted here are dominated by the lattice
results, which for these matrix elements are still in an early phase and the systematic uncertainties
are not fully explored yet. Still the error on the final result is already good (below ten percent),
and there is room for a large improvement which is expected in the near future. Note that when the uncertainties decrease sufficiently for results to become sensitive to isospin breaking effects, new couplings will appear 
in eq.~(\ref{eq:lagN}). These could in principle be extracted from lattice simulations by studying the explicit quark
mass dependence of the matrix element. In this regime the experimental value of the 
isovector coupling $g_A$ cannot be used anymore because of different isospin breaking corrections
to charged versus neutral currents.

The numerical values of the couplings we get are not too far off those already in the literature (see e.g.~\cite{Agashe:2014kda}). 
However, because of the caveats in the relation of the deep inelastic scattering and hyperon data 
to the relevant matrix elements the uncertainties in those approaches 
are not under control. On the other hand the lattice uncertainties are expected to improve in
the near future, which would further improve the precision of the estimate performed with
the technique presented here.

The numerical coefficients in eq.~(\ref{eq:cpcn}) include the 
effect of running from the high scale $f_a$ (here fixed to $10^{12}$~GeV) to the matching scale $Q=2$~GeV, 
which we performed at the NLLO order (more details in appendix~\ref{app:Zrge}). 
The running effects are evident from the fact that the couplings to nucleons
depend on all quark couplings including charm, bottom and top, 
even though we took the corresponding spin content to vanish.
This effect has been neglected in previous analysis.  

Finally it is interesting to observe that there is a cancellation in the model independent part of the axion coupling to the neutron
in KSVZ-like models, where $c_q^0=0$, 
\begin{equation}
c_p^{\rm KSVZ}=-0.47(3)\,, \qquad 
c_n^{\rm KSVZ}=-0.02(3)\,, 
\end{equation}
the coupling to neutrons is suppressed with respect to the coupling
to protons by a factor ${\cal O}(10)$ at least, in fact this coupling still is compatible with 0.
The cancellation can be understood from the fact that, neglecting running and sea quark contributions
\begin{equation}
c_n \sim  \left \langle Q_a \cdot \left ( \begin{array}{cc} \Delta d & 0 \\ 0 & \Delta u \end{array} \right) \right\rangle \propto m_d \Delta d+m_u \Delta u\,,
\end{equation}
and the down-quark spin content of the neutron $\Delta u$ is approximately $\Delta u\approx -2\Delta d$,  \emph{i.e.} the ratio $m_u/m_d$ is accidentally close to the ratio between the number of up over down valence quarks in the neutron. This cancellation may have important implications on axion detection and astrophysical bounds.

In models with $c_q^0\neq0$ both the couplings to proton and neutron can be large,
for example for the DFSZ axion models, where $c_{u,c,t}^0=\frac13\sin^2\beta=\frac13-c_{d,s,b}^0$ at the scale $Q\simeq f_a$,
we get
\begin{equation}
c_p^{\rm DFSZ}=-0.617+0.435\sin^2\beta \pm 0.025\,, \qquad 
c_n^{\rm DFSZ}=0.254-0.414\sin^2\beta \pm 0.025\,. 
\end{equation}
A cancellation in the coupling to neutrons is still possible for special values of $\tan\beta$.

\section{The hot axion: finite temperature results}  \label{sec:Tne0}
We now turn to discuss the properties of the QCD axion at finite temperature. The temperature dependence
of the axion potential and its mass are important in the early Universe because they control the relic abundance of axions today (for a  review see e.g. \cite{Sikivie:2006ni}). The most model independent mechanism of axion production in the early universe, the  misalignment mechanism \cite{Abbott:1982af,Preskill:1982cy,Dine:1982ah}, is almost completely determined by the shape of the axion potential at finite temperature and its zero temperature mass. Additionally, extra contributions, such as string and domain walls can also be present
if the PQ preserving phase is restored after inflation, and might be the dominant source of dark matter \cite{Sikivie:1982qv,Vilenkin:1982ks,Vilenkin:1984ib,
Davis:1986xc,Bennett:1987vf,Dabholkar:1989ju,Vincent:1996rb}. Their contribution also depends on the finite temperature behavior
of the axion potential, although there are larger uncertainties  in this case coming from the details
of their evolution (for a recent numerical study see e.g. \cite{Kawasaki:2014sqa}).\footnote{Axion could also be produced thermally in the early universe, this population would be sub-dominant for the allowed values of $f_a$ \cite{Berezhiani:1992rk,Masso:2002np,Graf:2010tv,Salvio:2013iaa} 
but might leave a trace as dark radiation.}

One may naively think that, as the temperature is raised, our knowledge of axion properties gets 
better and better---after all the higher the temperature the more perturbative QCD gets. 
The opposite is instead true. In this section we show that, at the moment, 
the precision with which we know the axion potential worsens as the temperature is increased!

At low temperature this is simple to understand. Our high precision estimates at zero temperature rely
on chiral Lagrangians whose convergence degrades as the temperature approaches the critical temperature $T_c \simeq$160-170~MeV where QCD starts deconfining. At $T_c$ the chiral approach
is already out of control. 
Fortunately around the QCD cross-over region lattice computations are possible. The current precision is not yet competitive with our low temperature results but they are expected to improve soon. At higher
temperatures there are no lattice results available. For $T\gg T_c$ the dilute instanton gas approximation,
being a perturbative computation, is believed to give a reliable estimate of the axion potential. 
It is known however that finite temperature QCD converges fast only for very large temperatures, 
above ${\cal O}(10^6)$~GeV (see e.g.~\cite{Andersen:2011sf}). The situation is particularly bad 
for the instanton computation. The screening of  QCD charge causes an exponential sensitivity to 
quantum thermal loop effects. The resulting uncertainty on the axion mass and potential can 
easily be one order of magnitude or more! This is compatible with a recent lattice computation 
\cite{Borsanyi:2015cka}, performed without quarks, which found a high temperature axion mass 
differing from the instanton prediction at $T= 1$ GeV by a factor $\sim 10$. More recent
preliminary results from simulations with dynamical quarks \cite{Trunin:2015yda} 
seem to show an even bigger
disagreement, perhaps suggesting that at these temperatures even the form of the action is very different from the instanton prediction.

\subsection{Low temperatures}
For temperatures $T$ below $T_c$ axion properties can reliably be computed within finite temperature
chiral Lagrangians \cite{Gasser:1986vb,Gasser:1987ah}. Given the QCD mass gap in this regime temperature effects are exponentially suppressed. 

The computation of the axion mass is straightforward. 
Note that the temperature dependence can only come 
from the non local contributions that can feel the finite 
temperature. At one loop the axion mass only receives  
contribution from the local NLO couplings once rewritten 
in terms of the physical $m_\pi$ and $f_\pi$  \cite{Hansen:1990yg}. This means 
that the leading temperature dependence is completely 
determined by the temperature dependence of $m_\pi$ 
and $f_\pi$, and in particular is the same as that of the 
chiral condensate \cite{Gasser:1986vb,Gasser:1987ah,Hansen:1990yg}
\begin{align} \label{eq:maT}
\frac{m_a^2(T)}{m_a^2} = \frac{\chi_{top}(T)}{\chi_{top}} \stackrel{\rm NLO}= \frac{m_\pi^2(T) f_\pi^2(T)}{m_\pi^2 f_\pi^2}
& =\frac{\langle \bar q q \rangle_T}{\langle \bar q q\rangle}=1-\frac{3}{2} \frac{T^2}{f_\pi^2}\, J_1\hspace{-3pt}\left [ \frac{m_\pi^2}{T^2}\right ] \,,
\end{align}
where
\begin{align}
J_{n}[\xi] = \frac{1}{(n-1)!}\left ( -\frac{\partial}{\partial \xi} \right)^n J_0[\xi]\,,\qquad J_0[\xi]  \equiv -\frac{1}{\pi^2}\int_{0}^{\infty}   dq \,q^2 \log\left(1-e^{-\sqrt{q^2+\xi}} \right)\,.
\end{align}
The function $J_1(\xi)$ asymptotes to $\xi^{1/4}e^{-\sqrt\xi}/(2\pi)^{3/2}$ 
at large $\xi$ and to $1/12$ at small $\xi$.
Note that in the ratio $m^2_a(T)/m^2_a$ the dependence 
on the quark masses and the NLO couplings cancel out. 
This means that, at $T\ll T_c$, this ratio is known at a even
 better precision than the axion mass at zero
temperature itself.

Higher order corrections are small for all values of $T$ below $T_c$. 
There are also contributions from the heavier states that are not 
captured by the low energy Lagrangian. In principle these
 are exponentially suppressed by $e^{-m/T}$, 
where $m$ is the mass of the heavy state. 
However, because the ratio $m/T_c$ is not very large and 
a large number of states appear above $T_c$ there is a 
large effect at around $T_c$, where the chiral expansion ceases to reliably
describe QCD physics. 
An in depth discussion of such effects appears in \cite{Gerber:1988tt}
for the similar case of the chiral condensate.

The bottom line is that for $T\lesssim T_c$ eq.~(\ref{eq:maT}) is a very good approximation
for the temperature dependence of the axion mass. At some temperature close to $T_c$ eq.~(\ref{eq:maT}) suddenly ceases to be a good approximation and full non-perturbative 
QCD computations are required.

The leading finite temperature dependence of the full potential can easily be derived as well,
\begin{equation} \label{eq:VaT}
\frac{V(a;T)}{V(a)} =1+\frac32 \frac{ T^4}{f_\pi^2 m_\pi^2\bigl ({\textstyle\frac{a}{f_a}}\bigr ) }\, J_0\hspace{-4pt}\left [\frac{m_\pi^2\bigl ({\textstyle\frac{a}{f_a}}\bigr )}{T^2}\right ]\,.
\end{equation}
The temperature dependent axion mass, eq.~(\ref{eq:maT}), can also be derived from eq.~(\ref{eq:VaT}) by taking the second derivative with respect to the axion.
The fourth derivative provides the temperature correction to the self-coupling,
\begin{equation}
\frac{\lambda_a(T)}{\lambda_a}=1-\frac{3}{2}\frac{T^2}{f_\pi^2}
J_1\hspace{-3pt}\left [ \frac{m_\pi^2}{T^2}\right ]
+\frac{9}{2}\frac{m_\pi^2}{f_\pi^2}\frac{m_u m_d}{m_u^2-m_u m_d+m_d^2} 
J_2\hspace{-3pt}\left [ \frac{m_\pi^2}{T^2}\right ]\,.
\end{equation}

\subsection{High temperatures}
While the region around $T_c$ is clearly in the non-perturbative regime,
for $T\gg T_c$ QCD is expected to become perturbative. At large temperatures the axion
potential can thus be computed in perturbation theory, around 
the dilute instanton gas background, as described in \cite{Gross:1980br}. 
The point is that, at high temperatures large gauge
configurations, which would dominate at zero temperature because of the
larger gauge coupling, are exponentially suppressed because of Debye screening. 
This makes the instanton computation a sensible one.

The prediction for the axion potential is of the form 
$V^{inst}(a;T)=-f_a^2 m_a^2(T)\cos(a/f_a)$ where
\begin{equation} \label{eq:mainst}
f_a^2 m_a^2(T) \simeq 2 \int d \rho\, n(\rho,0) e^{-\frac{2\pi^2}{g_s^2} m_{D1}^2 \rho^2
+\dots 
}\,,
\end{equation}
the integral is over the instanton size $\rho$,  
$n(\rho,0)\propto m_u m_d e^{-8\pi^2/g_s^2}$ is the zero temperature instanton density, 
$m_{D1}^2=g_s^2 T^2 (1+n_f/6)$ is the Debye mass squared at LO, 
$n_f$ is the number of flavor degrees of freedom active at the temperature $T$,
 and the dots stand for smaller corrections (see \cite{Gross:1980br} for more details).
The functional dependence of eq.~(\ref{eq:mainst}) on temperature is approximately
a power law $T^{-\alpha}$ where $\alpha \approx 7+n_f/3+\dots$ is fixed by the QCD
beta function.

There is however a serious problem with this type of computation. The dilute instanton
gas approximation relies on finite temperature perturbative QCD. 
The latter really becomes perturbative only at very high temperatures 
$T\gtrsim 10^6$~GeV due to IR divergences of the thermal bath \cite{Linde:1980ts}. 
Further, due to the exponential dependence on quantum corrections, 
the axion mass convergence is even worse than many other observables. 
In fact the LO estimate of the Debye mass $m_{D1}^2$ receives ${\cal O}(1)$ 
corrections at the NLO for temperatures around few GeV \cite{Rebhan:1993az,Arnold:1995bh}. 
Non-perturbative
computations from lattice simulations \cite{Kajantie:1997pd,Philipsen:2000qv,Maezawa:2007fc} confirm the unreliability of the LO estimate.

Both lattice \cite{Maezawa:2007fc} and NLO \cite{Rebhan:1993az} results give a Debye mass $m_D \simeq 1.5\,m_{D1}$ where $m_{D1}$ is the leading perturbative result. Since the
Debye mass enters the exponent of eq.~(\ref{eq:mainst}) higher order effects can easily 
shift the axion mass at a given temperature by an order of magnitude or more.

Given the failure of perturbation theory in this regime of temperatures even the actual form
of eq.~(\ref{eq:mainst}) may be questioned and the full answer could differ from the 
semiclassical instanton computation even in the temperature dependence and in 
the shape of the potential. Because of this, direct computations from non-perturbative 
methods such as lattice QCD are highly welcome.

Recently several computations of the temperature dependence of the topological susceptibility
for pure SU(3) Yang-Mills  appeared \cite{Berkowitz:2015aua,Borsanyi:2015cka}. While computations in this theory 
cannot be used for the QCD axion\footnote{Note that quarkless QCD differs from real QCD both
quantitatively (e.g. $\chi(0)^{1/4}=181~$MeV $vs$ $\chi(0)^{1/4}=75.5~$MeV, 
$T_c\simeq300$~MeV $vs$ $T_c\simeq160$~MeV) and qualitatively 
(the former undergoes a first order phase transition across 
$T_c$ while the latter only a crossover).}, they 
are useful to test the instanton result. 
In particular in \cite{Borsanyi:2015cka} an explicit comparison was made in the 
interval of temperatures $T/T_c \in [0.9,4.0]$. The results for the temperature dependence 
and the quartic derivative of the potential are compatible with those predicted by the 
instanton approximation, however the overall size of the topological susceptibility was found
one order of magnitude bigger. While the size of the discrepancy seem to be compatible
with a simple rescaling of the Debye mass, it goes in the opposite direction with respect
to the one suggested by higher order effects, preferring a smaller value for $m_D \simeq 0.5 m_{D1}$. This fact betrays a deeper modification of eq.~(\ref{eq:mainst}) than a simple 
renormalization of $m_D$.
\begin{figure}[t!]
\centering
\includegraphics[scale=0.6]{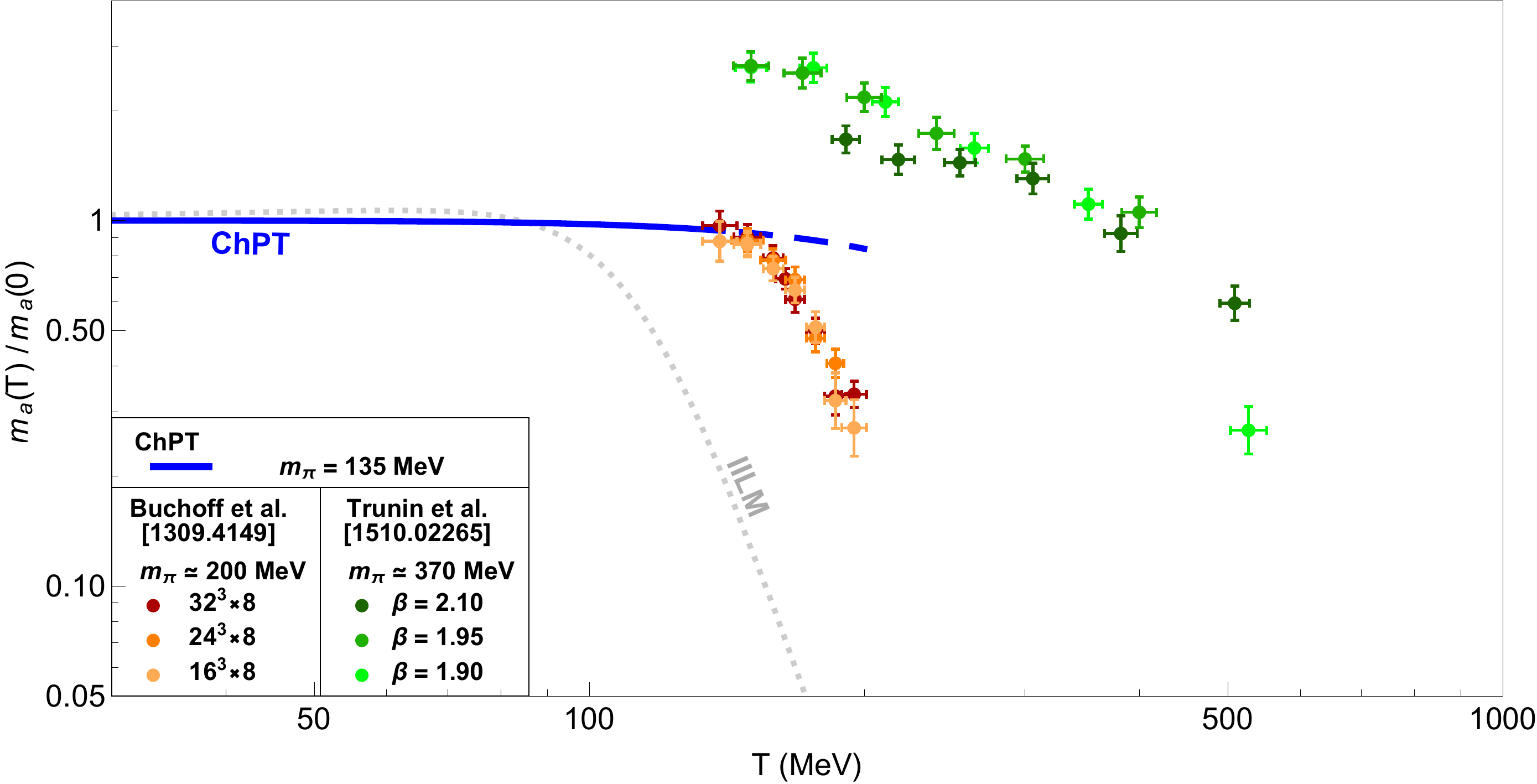}
\caption{The temperature dependent axion mass normalized to the zero temperature value (corresponding to the light quark mass values in each computation). In blue the prediction from
chiral Lagrangians. In different shades of red the lattice data from ref.~\cite{Buchoff:2013nra} for different lattice volumes, and in shades of green the preliminary lattice data from \cite{Trunin:2015yda} for different lattice spacings. The dotted grey curve shows the interacting instanton liquid model (IILM) result \cite{Wantz:2009mi}. \label{fig:maT}}
\end{figure}

Unfortunately no full studies for real QCD are available yet in the same range of temperatures.
Results across the crossover region, for $T\in[140,200]$~MeV, are available in 
\cite{Buchoff:2013nra}, 
which used light quark masses corresponding to $m_\pi\simeq 200$~MeV. 
Fig.~\ref{fig:maT} compares these results with the ChPT ones,  
with nice agreement around $T\sim 140$~MeV. The plot is in terms of the ratio $m_a(T)/m_a$, which at low temperatures weakens the quark mass dependence, as manifest in the ChPT computation. However, at high temperature this may not be true anymore.
For example the dilute instanton computation suggests $m_a^2(T)/m^2_a \propto (m_u+m_d)\propto m_\pi^2$, 
which implies that the slope across the crossover region may be very sensitive to the value
of the light quark masses.  In future lattice computations it is thus crucial to use physical quark masses, or at least to perform a reliable extrapolation to the physical point.

Additionally, while the volume dependence of the results in \cite{Buchoff:2013nra} 
seems to be under control, the lattice spacing used was rather coarse ($a>0.125$~fm) 
and furthermore not constant with the temperature.
Should the strong dependence 
on the lattice spacing observed in \cite{Borsanyi:2015cka} 
be also present in full QCD lattice simulations
a continuum limit extrapolation would become compulsory.

More recently, new preliminary lattice results appeared in \cite{Trunin:2015yda}, 
for a wider range of temperatures between 150 and 500~MeV. This analysis was performed
with 4 dynamical flavors, including the charm quark, 
but with heavier light quark masses, corresponding
to $m_\pi\simeq 370$~MeV. These results are also shown in fig.~\ref{fig:maT}, and suggest that $\chi(T)$ decreases with temperature 
much more slowly than in the quarkless case, 
in clear contradiction to the instanton calculation.
The analysis also includes different lattice spacing,
showing strong discretization effects. 
Given the strong dependence on the lattice spacing observed,
and the large pion mass employed, a proper analysis 
of the data is required before a direct comparison with 
the other results can be performed. In particular, the low temperature lattice points exceed the zero temperature chiral perturbation theory result (given their pion mass), which is presumably a consequence of the finite lattice spacing.

If the results for the temperature slope in \cite{Trunin:2015yda} are confirmed
in the continuum limit and for physical quark masses, it would
imply a temperature dependence for the topological susceptibility
($\chi(T)\sim T^{-2}$) departing strongly from the one predicted  
by instanton computations. As we will see in the next section 
this could have dramatic consequences 
in the computation of the axion relic abundance.

For completeness in fig.~\ref{fig:maT} we also show the result of \cite{Wantz:2009mi} obtained from an instanton-inspired model, which is sometimes used as input in the computation of the axion relic abundance. 
Although the dependence at low temperatures explicitly violates low-energy theorems 
the behaviour at higher temperature is similar to the lattice data by
\cite{Buchoff:2013nra}, although with a quite different $T_c$.

\subsection{Implications for dark matter}

The amount of axion dark matter produced in the early Universe and its properties depend
on whether PQ symmetry is broken or not after inflation.
If the PQ symmetry is broken before inflation ($H_I \lesssim f_a$) and not restored during 
reheating ($T_{\rm max}\lesssim f_a$), after the Big Bang the axion field is uniformly constant 
 over the  observable Universe, $a(x) = \theta_0 f_a$.  
The evolution of the axion field, in particular of its zero mode, is described by the equation of motion
\begin{equation} \label{eq:equa}
\ddot{a} + 3 H \dot{a} + m_a^2\left(T\right) f_a \sin\left(\frac{a}{f_a}\right) =0\,.
\end{equation}
where we assumed that the shape of the axion potential is well described by the dilute
instanton gas approximation, i.e. cosine like. 
As the Universe cools, the Hubble parameter decreases
while the axion potential increases. When the pull from the latter becomes 
comparable to the Hubble friction, i.e. $m_a(T)\sim 3 H$, the axion field starts oscillating with
frequency $m_a$. This  typically happens at temperatures above $T_c$, around  the GeV scale,
depending on the value of $f_a$ and the temperature dependence of the axion mass. 
Soon after that the comoving number density 
$n_a=\langle m_a a^2\rangle$ becomes an adiabatic invariant and the axion behaves
as cold dark matter.
\begin{figure}[t!]
\centering
\includegraphics[scale=0.5]{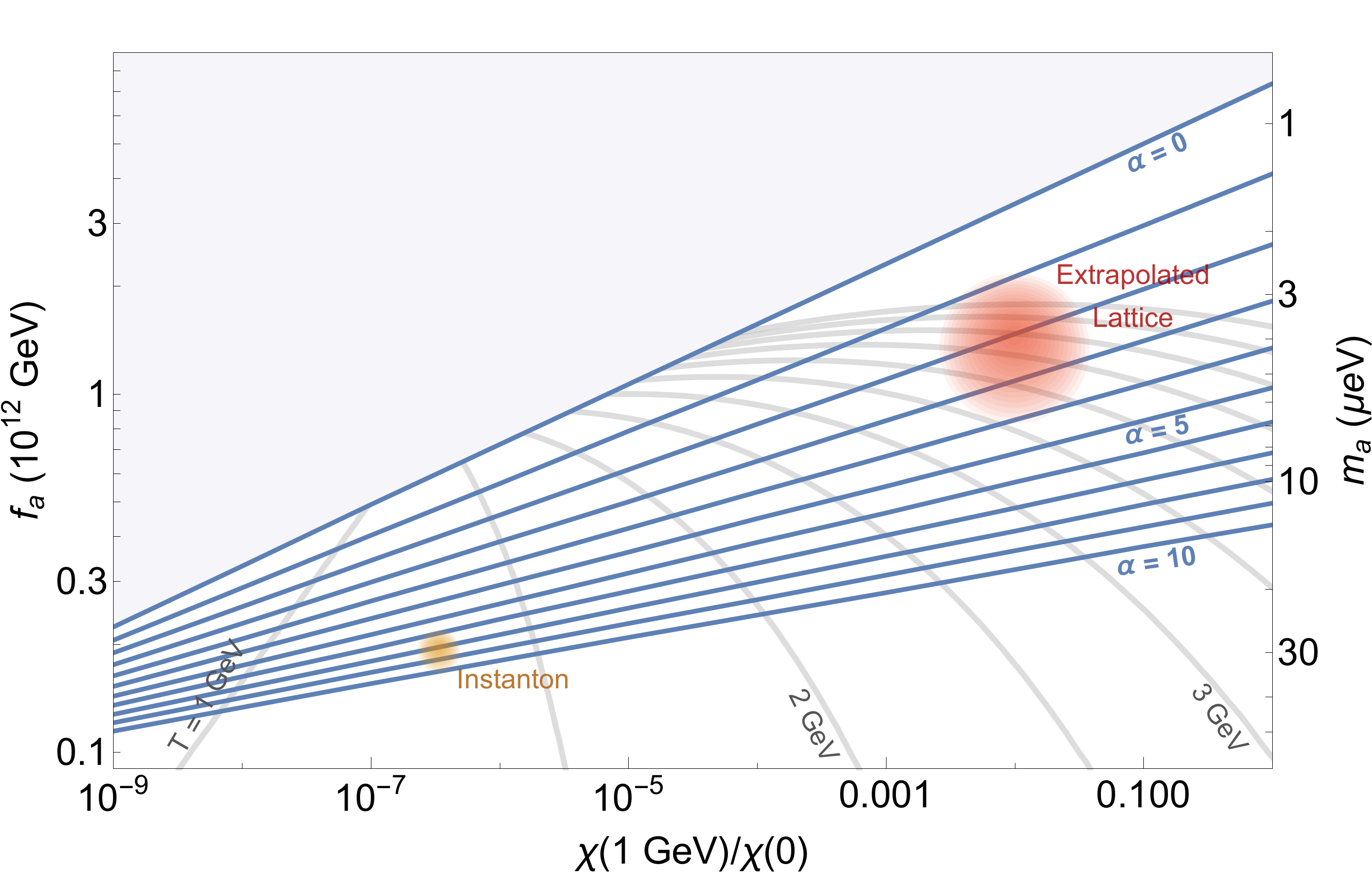}
\caption{ \label{fig:fachi} Values of $f_a$ such that the misalignment contribution
to the axion abundance matches the observed dark matter one for different choices
of the parameters of the axion mass dependence on temperature.
For definiteness the plot refers to the case where the PQ phase is restored after
the end of inflation (corresponding approximately to the choice $\theta_0=2.15$). 
The temperatures where the axion starts oscillating, 
i.e. satisfying the relation $m_a(T)=3H(T)$, are also shown. The two points corresponding
to the dilute instanton gas prediction and the recent preliminary lattice data are shown
for reference.}
\end{figure}

Alternatively PQ symmetry may be broken after inflation. In this case, immediately after the breaking the axion field finds itself randomly distributed over the whole range $\left[0,2\pi f_a \right]$. Such field configurations include strings which evolve with a complex dynamics, but are known to approach a scaling solution \cite{Bennett:1987vf}. At temperatures close to $T_c$, when the axion field starts rolling because of the QCD potential, domain walls also form. In phenomenologically viable models, the full field configuration, including strings and domain walls, eventually decays into
axions, whose abundance is affected by large uncertainties associated with the evolution
and decay of the topological defects.
Independently of this evolution there is a misalignment contribution to the dark matter relic density from axion modes with very close to zero momentum. The calculation of this is the same as for the case where inflation happens after PQ breaking, except that the relic density must be averaged over all possible values of  $\theta_0$. While the misalignment contribution gives only a part of the full abundance, it can still be used to give an upper bound to $f_a$ in this scenario.
\begin{figure}[t!]
\centering
\includegraphics[scale=0.5]{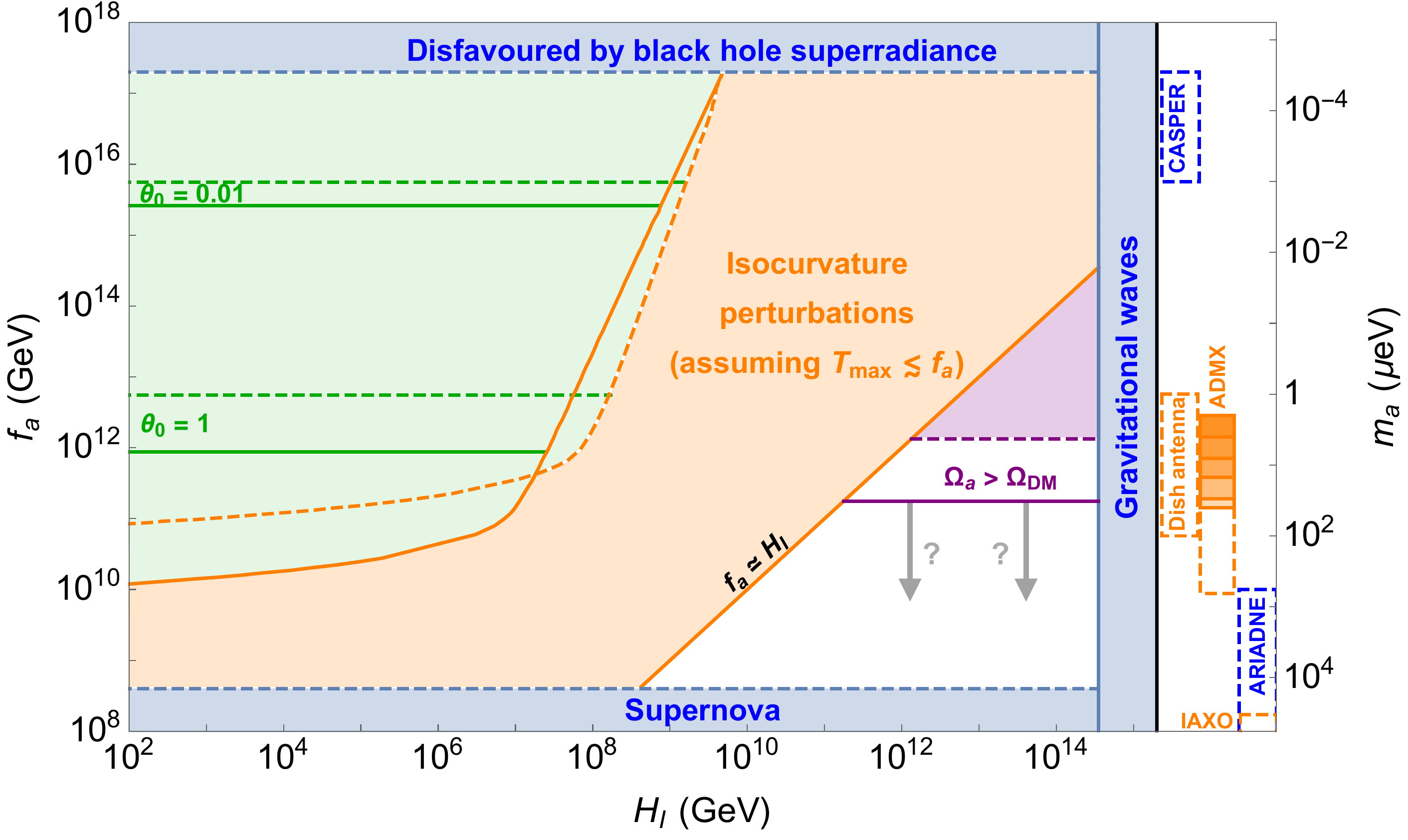}
\caption{The axion parameter space as a function of the axion decay constant and the Hubble parameter during inflation.  The bounds are shown for the two choices for the axion mass parametrization suggested  
by instanton computations (continuous lines) and by preliminary lattice results (dashed lines), corresponding to the labeled points in fig.~\ref{fig:fachi}. In the green shaded region the misalignment axion relic density can make up the entire dark matter abundance, and the isocurvature limits are obtained assuming that this is the case. In the white region the axion misalignment population can only be a sub-dominant component of dark matter. The region where PQ symmetry is restored after inflation does not include the contributions from topological defects, the lines thus only represent conservative upper bounds to the value of $f_a$.
Ongoing (solid) and proposed (dashed empty) experiments testing the available axion parameter
space are represented on the right side. \label{fig:fahi} } 
\end{figure}
%

The current axion abundance from misalignment, assuming standard cosmological evolution, is given by
\begin{equation}
\Omega_a=\frac{86}{33} \frac{\Omega_\gamma}{T_\gamma }  \frac{n_a^\star}{s^\star}m_a\,,
\end{equation}
where $\Omega_\gamma$ and $T_\gamma$ are the current photon abundance and 
temperature respectively and $s^\star$ and $n_a^\star$ are the entropy density and the
average axion number density computed at any moment in time $t_\star$ sufficiently after
the axion starts oscillating such that $n_a^\star/s^\star$ is constant.
The latter quantity can be obtained by solving eq.~(\ref{eq:equa}) and depends on 
1) the QCD energy and entropy density around $T_c$, 2) the initial condition for the axion field $\theta_0$, and 3) the temperature dependence of the axion mass and potential.
The first is reasonably well known from perturbative methods and lattice simulations (see e.g.~\cite{Philipsen:2012nu,Borsanyi:2013bia}). 
The initial value $\theta_0$ is a free parameter in the first scenario, where
the PQ transition happen before inflation---since in this case $\theta_0$ can be chosen
in the whole interval $[0,2\pi]$ only an upper bound to $\Omega_a$ can be obtained in this case.
In the scenario where the PQ phase is instead restored after inflation $n_a^\star$ is obtained
by averaging over all $\theta_0$, which numerically corresponds to choosing\footnote{The effective
$\theta_0$ corresponding to the average is somewhat bigger than $\langle \theta^2 \rangle=\pi^2/3$ because of anharmonicities of the axion potential. } 
$\theta_0\simeq 2.1$. Since $\theta_0$ is fixed, $\Omega_a$ is completely determined as a function of $f_a$ in this case. 
At the moment the biggest uncertainty on the misalignment contribution to 
$\Omega_a$ comes from our knowledge of $m_a(T)$.
Assuming that $m_a(T)$ can be approximated by the power law 
$$
m_a^2(T)=m_a^2(1~{\rm GeV}) \left(\frac{{\rm GeV}}T\right)^\alpha=
m_a^2\, \frac{\chi(1~{\rm GeV})}{\chi(0)} \left(\frac{{\rm GeV}}T\right)^\alpha \,,
$$
around the temperatures where the axion starts oscillating, eq.~(\ref{eq:equa}) 
can easily be integrated numerically. In fig.~\ref{fig:fachi}
 we plot the values of $f_a$ that would reproduce the correct dark matter 
abundance for different choices of $\chi(T)/\chi(0)$ and $\alpha$ in the scenario
where $\theta_0$ is integrated over. We also show two representative points
with parameters ($\alpha\approx 8$, $\chi$(1~GeV)$/\chi(0)\approx$~few~$10^{-7}$) and  ($\alpha\approx 2$, $\chi$(1~GeV)$/\chi(0)\approx 10^{-2}$) corresponding respectively
to the expected behavior from instanton computations and to the suggested one from 
the preliminary lattice data in \cite{Trunin:2015yda}.
The figure also shows the corresponding temperature at which the axion starts
oscillating, here defined by the condition $m_a(T)=3H(T)$.

Notice that for large values of $\alpha$, as predicted by instanton computations,
the sensitivity to the overall size of the axion mass at fixed temperature 
($\chi$(1~GeV)$/\chi$(0)) is weak. However if the slope of the axion mass
with the temperature is much smaller, as suggested by the results in \cite{Trunin:2015yda}, 
then the corresponding value of $f_a$ required to give the correct relic abundance
can even be larger by an order of magnitude (note also that in this case 
the temperature at which the axion starts oscillating would be higher, around 4$\div$5~GeV).
The difference between the two cases could be taken as an estimate 
of the current uncertainty on this type of computation.
More accurate lattice results would be very welcome to assess the actual temperature
dependence of the axion mass and potential.

To show the impact of this uncertainty on the viable axion parameter space and
the experiments probing it, in fig.~\ref{fig:fahi} we plot the various constraints as a function of the Hubble scale during inflation and the axion decay constant. Limits that depend on the
temperature dependence of the axion mass are shown for the instanton and lattice inspired forms (solid and dashed lines respectively), corresponding to the labeled points in fig.~\ref{fig:fachi}.
On the right side of the plot we also show the values of $f_a$ that will be probed
by ongoing experiments (solid) and those that could be probed
by proposed experiments (dashed empty). Orange colors are used for experiments using
the axion coupling to photons, blue for the others. Experiments in
the last column (IAXO and ARIADNE) do not rely on the axion being dark matter.
The boundary of the allowed axion parameter space is constrained by the CMB limits 
on tensor modes \cite{Ade:2015lrj}, supernova SN1985 and other 
astrophysical bounds 
including black-hole superradiance. 

When the PQ preserving phase is not restored after inflation (i.e. when both the Hubble parameter
during inflation $H_I$ and the maximum temperature after inflation $T_{max}$ are smaller
than the PQ scale)  the axion abundance can match the observed dark matter one
for a large range of values of $f_a$ and $H_I$ by varying the initial axion value $\theta_0$.
In this case isocurvature bounds \cite{Linde:1985yf} (see e.g.~\cite{Hamann:2009yf} 
for a recent discussion) constrain $H_I$ from above. At small $f_a$ obtaining the correct relic abundance requires $\theta_0$ to be close to $\pi$, where the potential is flat, so the the axion begins oscillating at relatively late times. In the limit $\theta_0\to \pi$ the axion energy density diverges.
Given the sensitivity of $\Omega_a$ to $\theta_0$ in this regime,
isocurvatures are enhanced by $1/(\pi-\theta_0)$ and 
the bound on $H_I$ is thus strengthened by a factor $\pi-\theta_0$.\footnote{This constraint guarantees that we are consistently working in a regime where quantum fluctuations during inflation are much smaller than the distance of the average value of $\theta_0$ from the top of the potential.} 
Meanwhile, the axion decay constant is bounded from above by black-hole superradiance. For smaller values of $f_a$ axion misalignment can only explain part of the dark matter abundance. In fig.~\ref{fig:fahi} we show the value of $f_a$ required  to explain $\Omega_{DM}$ when $\theta_0 = 1$ and $\theta_0 = 0.01$ for the two reference
values of the axion mass temperature parameters.


If the PQ phase is instead restored after inflation, e.g. for high scale inflation models,
$\theta_0$ is not a free parameter anymore. In this case only one value of $f_a$ will
reproduce the correct dark matter abundance. Given our ignorance about the 
contributions from topological defect we can use the misalignment computation
to give an upper bound on $f_a$. This is shown on the bottom-right side of the plot,
again for the two reference models, as before. 
Contributions from higher-modes and topological defects are likely to make such
bound stronger by shifting the forbidden region downwards.
Note that while the instanton behavior for the temperature dependence of the axion mass
would point to axion masses outside the range which will be probed by
ADMX (at least in the current version of the experiment), if the lattice behavior will
be confirmed the mass window which will be probed would look much more promising.

\section{Conclusions}  \label{sec:concl}
We showed that several QCD axion properties, despite being determined by
non-perturbative QCD dynamics, can be computed reliably with high accuracy. 
In particular we computed higher order corrections to the axion mass, its self-coupling, the coupling to photons, the full potential and the domain-wall tension, providing estimates
for these quantities with percent accuracy. 
We also showed how lattice data can be used to extract the axion coupling to matter (nucleons) 
reliably providing estimates with better than 10\% precision. 
These results are important both experimentally, to assess the actual axion parameter
space probed and to design new experiments, and theoretically, since
in the case of a discovery they would help determining the underlying theory behind
the PQ breaking scale.

We also study the dependence of the axion mass and potential on the temperature,
which affects the axion relic abundance today. While at low temperature such information
can be extracted  accurately using chiral Lagrangians at temperatures close to
the QCD crossover and above perturbative methods fail. We also point out that 
instanton computations, which are believed to become reliable at least when QCD becomes
perturbative have serious convergence problems, making them unreliable in the whole
region of interest. Recent lattice results seem indeed to suggest large deviations from
the instanton estimates. We studied the impact that this uncertainty has on
the computation of the axion relic abundance and the constraints on the axion parameter
space. More dedicated non-perturbative computations are therefore required to reliably determine
the axion relic abundance.

\section*{Acknowledgments}
This work is supported in part by the ERC Advanced Grant no.267985 (DaMeSyFla).

\appendix

\section{Input parameters and conventions} \label{sec:app1}
For convenience  in table~\ref{tab:numval} we report the values of the parameters used in this work. When uncertainties are not quoted it means that their effect was negligible and they have not been used.
  
In the following we discuss in more in details the origin of some of these values.

\begin{table}[t]
\centering
\begin{tabular}{||c|c||c|c||}
\hline  \hline
$z$ & 0.48(3) &  $ \bar l_3$ & 3(1)  \\ \hline 
$r$ & 27.4(1) & $\bar l_4$ & 4.0(3)  \\ \hline
$m_\pi$ & 134.98 &  $l_7$ & 0.007(4) \\ \hline 
$m_K$ & 498 & $L_7^r$ & $-$0.0003(1) \\ \hline
$m_\eta$ & 548 & $L_8^r$ & 0.00055(17) \\ \hline
$f_\pi$ & 92.2 & $g_A$ & 1.2723(23)\\ \hline
$f_\eta/f_\pi$ & 1.3(1) & $\Delta u+\Delta d$ & 0.52(5) \\ \hline 
$\Gamma_{\pi\gamma\gamma}$ & $5.16(18)\, 10^{-4}$ & $\Delta s$ & $-$0.026(4)  \\ \hline
$\Gamma_{\eta\gamma\gamma}$ & $7.63(16) \,10^{-6}$ & $\Delta c$ & 0.000(4)  \\ \hline \hline
\end{tabular}
\caption{\label{tab:numval} Numerical input values used in the computations. 
Dimensionful quantities are given in MeV. The values of scale dependent low-energy constants are given at the scale $\bar \mu=770$~MeV, while the scale dependent proton spin content $\Delta q$ are given at $Q=2$~GeV.}
\end{table}

\subsection*{Quark masses}
The value of $z=m_u/m_d$ has been extracted from the following lattice estimates:
\begin{equation}
z=\left \{ \begin{array}{l l}
0.52(2) &  \text{\cite{Horsley:2015eaa}} \\
0.50(2)(3) & \text{\cite{deDivitiis:2013xla}} \\
0.451(4)(8)(12) & \text{\cite{Basak:2015lla}}
\end{array}  
\right.
\end{equation}
which use different techniques, fermion formulations, etc. 
In \cite{Sanfilippo:2015era} the extra preliminary result $z=0.49(1)(1)$ is also 
quoted, which agrees with the results above.
Some results are still preliminary and the study 
of systematics may not be complete. Indeed the spread from the central values
is somewhat bigger than the quoted uncertainties.
Averaging the results above we get $z=0.48(1)$.
Waiting for more complete results and a more systematic
study of all uncertainties we used a more conservative 
error, $z=0.48(3)$, which better captures the spread
between the different computations.

Axion properties have a much weaker dependence on the strange quark mass
which only enter at higher orders. For definiteness we used the value of the ratio
\begin{equation}
r\equiv \frac{2 m_s}{m_u+m_d}=27.4(1)\,,
\end{equation}
from  \cite{Sanfilippo:2015era}.

 \subsection*{ChPT low energy constants}
For the value of the pion decay constant we used the PDG \cite{Agashe:2014kda} value:
\begin{equation}
f_{\pi}=92.21(14)~{\rm MeV}\,,
\end{equation} 
which is free from the leading EM corrections present in the leptonic decays
used for the estimates.

Following \cite{Gasser:1984gg} the ratio $f_\eta/f_\pi$ can be related to $f_K/f_\pi$, 
whose value is very well known, up to higher order corrections. Assuming the usual 30\% 
uncertainty on the $SU(3)$ chiral estimates we get $f_\eta/f_\pi=1.3(1)$.

For the NLO low energy couplings we used the usual conventions of \cite{Gasser:1983yg,Gasser:1984gg}. As described in the main text we used
the matching of the 3 and 2 flavor Lagrangians to estimate the SU(2) couplings
from the SU(3) ones. In particular we only need the values of $L_{7,8}^r$,
which we took as
\begin{equation}
L_7^r\equiv L_7^r(\bar \mu)=-0.3(1)\cdot 10^{-3}\,,\qquad
L_8^r\equiv L_8^r(\bar \mu)=0.55(17)\cdot 10^{-3}\,,
\end{equation}
computed at the scale $\bar\mu=770$~MeV. The first number has been extracted
from the fit in \cite{Bijnens:2014lea} using the constraints for $L_4^r$ in \cite{Aoki:2013ldr}. 
The second from \cite{Aoki:2013ldr}. A 30\% intrinsic uncertainty from 
higher order 3-flavor corrections has been added. This intrinsic uncertainty is not
present for the 2-flavor constants where higher order corrections are much smaller.

In the main text we used the values 
\begin{align}
\bar l_3& =3(1)\,, \qquad 
l_3^r(\bar \mu) = -\frac{1}{64\pi^2} \left(  \bar l_3 +\log \left(\frac{m_\pi^2}{\bar\mu^2} \right)\right)\,,
\nonumber \\
\bar l_4& =4.0(3)\,, \qquad 
l_4^r(\bar \mu) = \frac{1}{16\pi^2} \left(  \bar l_4 +\log \left(\frac{m_\pi^2}{\bar\mu^2} \right)\right)\,,
\nonumber
\end{align}
extracted from 3-flavor simulations in \cite{Aoki:2013ldr}.

From the values above and using the matching in \cite{Gasser:1984gg}  
between the 2 and the 3 flavor theories  we can also extract:
\begin{equation}
l_7=7(4)\, 10^{-3}\,, \qquad h_1^r-h_3^r-l_4^r=-0.0048(14)\,.
\end{equation}

Preliminary results using estimates from lattice QCD simulations \cite{Mawhinney:lattice15} give
$\bar l_3=2.97(19)(14)$, $\bar l_4=3.90(8)(14)$, $l_7=0.0066(54)$ and 
$L_8^r=0.51(4)(12)\,10^{-3}$. The new results in \cite{Boyle:2015exm}
using partially quenched simulations give
$\bar l_3=2.81(19)(45)$, $\bar l_4=4.02(8)(24)$ and $l_7=0.0065(38)(2)$. All these results are in agreement with the numbers
used here.

\subsection*{Proton spin content}
While the axial charge, which is equivalent to the isovector spin content of the proton, is
very well known (see discussion around eq.~(\ref{eq:gAnum})) the isosinglet components
are less known.

To estimate $g^{ud}=\Delta u+\Delta d$ 
we use the results in \cite{QCDSF:2011aa,Engelhardt:2012gd,
Abdel-Rehim:2013wlz,Bhattacharya:2015gma,Abdel-Rehim:2015owa,Abdel-Rehim:2015lha}. 
In particular we used \cite{Abdel-Rehim:2015owa}, whose value for $g_A=1.242(57)$
is compatible with the experimental one,  to estimate the connected contribution
to  $g^{ud}$. For the disconnected contribution, which is much more difficult to
simulate, we averaged the results in \cite{Abdel-Rehim:2013wlz,Bhattacharya:2015gma,Abdel-Rehim:2015lha} increasing the error to accommodate the spread in central values,
which may be due to different systematics. 
Combining the results we get
\begin{equation}
g^{ud}_{conn.}+g^{ud}_{disc.}=0.611(48)-0.090(20)=0.52(5)\,.
\end{equation} 
All the results  provided here are in the $\overline {\rm MS}$ scheme at the reference scale $Q=2$~GeV.

The strange spin contribution only have the disconnected contribution, which we extract
averaging the results in \cite{QCDSF:2011aa,Engelhardt:2012gd,
Abdel-Rehim:2013wlz,Bhattacharya:2015gma,Abdel-Rehim:2015lha}
\begin{equation}
g^s=\Delta s=-0.026(4)\,.
\end{equation} 
All the results mostly agree with each others but they are still preliminary 
or use heavy quark masses or coarse lattice spacing or only two dynamical quarks. 
For this reason the estimate of the systematic uncertainties is not yet 
complete and further studies are required.


Finally \cite{Abdel-Rehim:2013wlz} also explored the charm spin contribution.
They could not see a signal and thus their results can only be used to put an upper bound
which we extracted as in table~\ref{tab:numval}.

\section{Renormalization of axial couplings} \label{app:Zrge}
While anomalous dimensions of conserved currents vanish it is not true for anomalous currents.
This means that the axion coupling to the singlet component of the axial current is scale dependent:
\begin{align}
&\frac{\partial_\mu a}{2 f_a} \sum_q  c_q j^\mu_q
 =\frac{\partial_\mu a}{2 f_a} \left[
 \sum_q  \left (c_q-\frac{\sum_{q'} c_{q'}}{n_f} \right )j_q^\mu
 +\frac{\sum_{q'} c_{q'}}{n_f}j^\mu_{\Sigma q} \right]  \\
& \quad \to \quad   \frac{\partial_\mu a}{2 f_a}  \left[ \sum_q \left (c_q-\frac{\sum_{q'} c_{q'}}{n_f}  \right) j_q^\mu
 +Z_0(Q)\frac{\sum_{q'} c_{q'}}{n_f}j^\mu_{\Sigma q} \right]
\end{align}
where $Z_0(Q)$ is the renormalization of the singlet axial current $j^\mu_{\Sigma q}$. 
It is important to note that $j^\mu_{\Sigma q}$ only renormalizes multiplicatively, this is not true
for the coupling to the gluon operator ($G \tilde G$) which mixes at one-loop with $\partial_\mu j^\mu_{\Sigma q}$ after
renormalization (see e.g.~\cite{Altarelli:1988nr}).  

The anomalous dimension of $j^\mu_{\Sigma q}$ starts only at 2-loops and is known up to 3-loops in QCD \cite{Kodaira:1979pa,Larin:1993tq}
\begin{equation} \label{eq:Zgamma}
\frac{\partial \log Z_0(Q)}{\partial \log Q^2}=\gamma_A=\frac{n_f}{2}\left( \frac{\alpha_s}{\pi}\right)^2+n_f\frac{177-2n_f}{72} \left ( \frac{\alpha_s}{\pi}\right)^3+\dots\,.
\end{equation}
The evolution of the couplings $c_q(Q)$ can thus be written as
\begin{equation}
c_q(Q)=c_q( Q_0)+\left ( \frac{Z_0(Q)}{Z_0( Q_0)}-1\right)\frac{\langle c_q \rangle_{n_f}}{n_f} \,,
\end{equation}
where we used the short hand notation $\langle \cdot \rangle_{n_f}$ for the sum of $q$ over $n_f$ flavors. 
Iterating the running between the high scale $f_a$ and the low scale $Q=2$~GeV across the 
bottom and top mass thresholds we can finally write the relation between the low energy couplings $c_q(Q)$ and the high energy ones $c_q=c_q(f_a)$:
\begin{align} \label{eq:cqfZ}
c_t(m_t) & =c_t+\left ( \frac{Z_0(m_t)}{Z_0(f_a)}-1\right)\frac{\langle c_q \rangle_{6}}{6} \,, \nonumber \\
c_b(m_b) & =c_b+\left ( \frac{Z_0(m_b)}{Z_0(m_t)}-1\right)\frac{\langle c_q \rangle_{5}}{5}
+ \frac{Z_0(m_b)}{Z_0(m_t)}\left ( \frac{Z_0(m_t)}{Z_0(f_a)}-1\right)\frac{\langle c_q \rangle_{6}}{6} \,, \nonumber \\
c_{q=u,d,s,c}(Q)&=c_q+\left ( \frac{Z_0(Q)}{Z_0(m_b)}-1\right)\frac{\langle c_q \rangle_{4}}{4}+
 \frac{Z_0(Q)}{Z_0(m_b)}\left ( \frac{Z_0(m_b)}{Z_0(m_t)}-1\right)\frac{\langle c_q \rangle_{5}}{5}
\nonumber \\
& \quad + \frac{Z_0(Q)}{Z_0(m_t)}\left ( \frac{Z_0(m_t)}{Z_0(f_a)}-1\right)\frac{\langle c_q \rangle_{6}}{6} \,,
\end{align}
where at each mass threshold we matched the couplings at LO. In eq.~(\ref{eq:cqfZ}) we can recognize
the contributions from the running from $f_a$ to $m_t$ with 6 flavors, from $m_t$ to $m_b$ with 5 flavors
and the one down to $Q$ with 4 flavors.

The value for $Z_0(Q)$ can be computed from eq.~(\ref{eq:Zgamma}), at LLO the solution is simply
\begin{equation}
Z_{0}(Q)=Z_0(Q_0)\ e^{-\frac{6n_f}{33-2n_f}\frac{\alpha_s(Q)-\alpha_s(Q_0)}{\pi}}\,.
\end{equation}

At NLLO the numerical values at the relevant mass scales are
\begin{equation}
 Z_0(10^{12}~{\rm GeV})=0.984\,,\quad Z_0(m_t)=0.939(3) \,,
 \quad Z_0(m_b)=0.888(15)\,,\quad Z_0(2~{\rm GeV})=0.863(24)\,,
 \end{equation} 
where the error is estimated by the difference with the LLO which should capture
the order of magnitude of the 1-loop thresholds not included in the computation.
For the computation above we used the $\overline{\rm MS}$ values of the quark masses,
i.e. $m_t(m_t)=164$~GeV and $m_b(m_b)=4.2$~GeV.
The dependence of $Z_0(f_a)$ on the actual value of $f_a$ is very mild, shifting 
$Z_0(f_a)$ by less than $\pm0.5$\% for $f_a=10^{12\pm3}$~GeV.

Note that DFSZ models at high energy can be written so that the axion couples
only through the quark mass matrix. In this case no running effect should be present
above the first SM mass threshold (at the top mass). Indeed in this models,
$\langle c_q \rangle_6=\langle c^0_q\rangle_6-{\rm tr}\,Q_a=0$ and the renormalization
effects from $f_a$ to $m_t$ cancel out.



\begin{thebibliography}{99}

\bibitem{Crewther:1979pi}
  R.~J.~Crewther, P.~Di Vecchia, G.~Veneziano and E.~Witten,
  ``Chiral Estimate of the Electric Dipole Moment of the Neutron in Quantum Chromodynamics,''
  Phys.\ Lett.\ B {\bf 88} (1979) 123
   [Phys.\ Lett.\ B {\bf 91} (1980) 487].

\bibitem{Afach:2015sja}
  S.~Afach {\it et al.},
  ``A Revised Experimental Upper Limit on the Electric Dipole Moment of the Neutron,''
  arXiv:1509.04411 [hep-ex].
  
  
\bibitem{Peccei:1977hh}
  R.~D.~Peccei and H.~R.~Quinn,
  ``CP Conservation in the Presence of Instantons,''
  Phys.\ Rev.\ Lett.\  {\bf 38} (1977) 1440.
  
\bibitem{Wilczek:1977pj}
  F.~Wilczek,
  ``Problem of Strong p and t Invariance in the Presence of Instantons,''
  Phys.\ Rev.\ Lett.\  {\bf 40} (1978) 279.
  
\bibitem{Weinberg:1977ma}
  S.~Weinberg,
  ``A New Light Boson?,''
  Phys.\ Rev.\ Lett.\  {\bf 40} (1978) 223.

\bibitem{Kim:1979if}
  J.~E.~Kim,
  ``Weak Interaction Singlet and Strong CP Invariance,''
  Phys.\ Rev.\ Lett.\  {\bf 43} (1979) 103.
  
\bibitem{Shifman:1979if}
  M.~A.~Shifman, A.~I.~Vainshtein and V.~I.~Zakharov,
  ``Can Confinement Ensure Natural CP Invariance of Strong Interactions?,''
  Nucl.\ Phys.\ B {\bf 166} (1980) 493.

\bibitem{Zhitnitsky:1980tq}
  A.~R.~Zhitnitsky,
  ``On Possible Suppression of the Axion Hadron Interactions. (In Russian),''
  Sov.\ J.\ Nucl.\ Phys.\  {\bf 31} (1980) 260
   [Yad.\ Fiz.\  {\bf 31} (1980) 497].
    
\bibitem{Dine:1981rt}
  M.~Dine, W.~Fischler and M.~Srednicki,
  ``A Simple Solution to the Strong CP Problem with a Harmless Axion,''
  Phys.\ Lett.\ B {\bf 104} (1981) 199.

\bibitem{Vafa:1984xg}
  C.~Vafa and E.~Witten,
  ``Parity Conservation in QCD,''
  Phys.\ Rev.\ Lett.\  {\bf 53} (1984) 535.

\bibitem{Raffelt:2006cw}
  G.~G.~Raffelt,
  ``Astrophysical axion bounds,''
  Lect.\ Notes Phys.\  {\bf 741} (2008) 51
  [hep-ph/0611350].
  
\bibitem{Arvanitaki:2009fg}
  A.~Arvanitaki, S.~Dimopoulos, S.~Dubovsky, N.~Kaloper and J.~March-Russell,
  ``String Axiverse,''
  Phys.\ Rev.\ D {\bf 81} (2010) 123530
  [arXiv:0905.4720 [hep-th]].
  
\bibitem{Arvanitaki:2010sy}
  A.~Arvanitaki and S.~Dubovsky,
  ``Exploring the String Axiverse with Precision Black Hole Physics,''
  Phys.\ Rev.\ D {\bf 83} (2011) 044026
  [arXiv:1004.3558 [hep-th]].
  
\bibitem{Arvanitaki:2014wva}
  A.~Arvanitaki, M.~Baryakhtar and X.~Huang,
  ``Discovering the QCD Axion with Black Holes and Gravitational Waves,''
  Phys.\ Rev.\ D {\bf 91} (2015) 8,  084011
  [arXiv:1411.2263 [hep-ph]].
  
\bibitem{Preskill:1982cy}
  J.~Preskill, M.~B.~Wise and F.~Wilczek,
  ``Cosmology of the Invisible Axion,''
  Phys.\ Lett.\ B {\bf 120} (1983) 127.

\bibitem{Abbott:1982af}
  L.~F.~Abbott and P.~Sikivie,
  ``A Cosmological Bound on the Invisible Axion,''
  Phys.\ Lett.\ B {\bf 120} (1983) 133.

\bibitem{Dine:1982ah}
  M.~Dine and W.~Fischler,
  ``The Not So Harmless Axion,''
  Phys.\ Lett.\ B {\bf 120} (1983) 137.
 
\bibitem{Asztalos:2009yp}
  S.~J.~Asztalos {\it et al.} [ADMX Collaboration],
  ``A SQUID-based microwave cavity search for dark-matter axions,''
  Phys.\ Rev.\ Lett.\  {\bf 104} (2010) 041301
  [arXiv:0910.5914 [astro-ph.CO]].
 
\bibitem{Armengaud:2014gea}
  E.~Armengaud {\it et al.},
  ``Conceptual Design of the International Axion Observatory (IAXO),''
  JINST {\bf 9} (2014) T05002
  [arXiv:1401.3233 [physics.ins-det]].

\bibitem{Horns:2012jf}
  D.~Horns, J.~Jaeckel, A.~Lindner, A.~Lobanov, J.~Redondo and A.~Ringwald,
  JCAP {\bf 1304} (2013) 016
  [arXiv:1212.2970 [hep-ph]].

\bibitem{Budker:2013hfa}
  D.~Budker, P.~W.~Graham, M.~Ledbetter, S.~Rajendran and A.~Sushkov,
  ``Proposal for a Cosmic Axion Spin Precession Experiment (CASPEr),''
  Phys.\ Rev.\ X {\bf 4} (2014) 2,  021030
  [arXiv:1306.6089 [hep-ph]].
 
\bibitem{Arvanitaki:2014dfa}
  A.~Arvanitaki and A.~A.~Geraci,
  ``Resonantly Detecting Axion-Mediated Forces with Nuclear Magnetic Resonance,''
  Phys.\ Rev.\ Lett.\  {\bf 113} (2014) 16,  161801
  [arXiv:1403.1290 [hep-ph]].

\bibitem{Sikivie:1983ip}
  P.~Sikivie,
  ``Experimental Tests of the Invisible Axion,''
  Phys.\ Rev.\ Lett.\  {\bf 51} (1983) 1415
   [Phys.\ Rev.\ Lett.\  {\bf 52} (1984) 695].

\bibitem{Krauss:1985ub}
  L.~Krauss, J.~Moody, F.~Wilczek and D.~E.~Morris,
  ``Calculations for Cosmic Axion Detection,''
  Phys.\ Rev.\ Lett.\  {\bf 55} (1985) 1797.

\bibitem{Weinberg:1978kz}
  S.~Weinberg,
  ``Phenomenological Lagrangians,''
  Physica A {\bf 96} (1979) 327.

\bibitem{Gasser:1983yg}
  J.~Gasser and H.~Leutwyler,
  ``Chiral Perturbation Theory to One Loop,''
  Annals Phys.\  {\bf 158} (1984) 142.

\bibitem{Gasser:1984gg}
  J.~Gasser and H.~Leutwyler,
  ``Chiral Perturbation Theory: Expansions in the Mass of the Strange Quark,''
  Nucl.\ Phys.\ B {\bf 250} (1985) 465.
  
\bibitem{Buchoff:2013nra}
  M.~I.~Buchoff {\it et al.},
  ``QCD chiral transition, U(1)A symmetry and the dirac spectrum using domain wall fermions,''
  Phys.\ Rev.\ D {\bf 89} (2014) 5,  054514
  [arXiv:1309.4149 [hep-lat]]. 

\bibitem{Trunin:2015yda}
  A.~Trunin, F.~Burger, E.~M.~Ilgenfritz, M.~P.~Lombardo and M.~Muller-Preussker,
  ``Topological susceptibility from $N_f=2+1+1$ lattice QCD at nonzero temperature,''
  arXiv:1510.02265 [hep-lat].
  
\bibitem{Berkowitz:2015aua}
  E.~Berkowitz, M.~I.~Buchoff and E.~Rinaldi,
  ``Lattice QCD input for axion cosmology,''
  Phys.\ Rev.\ D {\bf 92} (2015) 3,  034507
  [arXiv:1505.07455 [hep-ph]].
    
 \bibitem{Borsanyi:2015cka}
  S.~Borsanyi {\it et al.},
  ``Axion cosmology, lattice QCD and the dilute instanton gas,''
  arXiv:1508.06917 [hep-lat].

\bibitem{DiVecchia:1980ve}
  P.~Di Vecchia and G.~Veneziano,
  ``Chiral Dynamics in the Large n Limit,''
  Nucl.\ Phys.\ B {\bf 171} (1980) 253.
  
\bibitem{Georgi:1986df}
  H.~Georgi, D.~B.~Kaplan and L.~Randall,
  ``Manifesting the Invisible Axion at Low-energies,''
  Phys.\ Lett.\ B {\bf 169} (1986) 73.
  
\bibitem{Ubaldi:2008nf}
  L.~Ubaldi,
  ``Effects of theta on the deuteron binding energy and the triple-alpha process,''
  Phys.\ Rev.\ D {\bf 81} (2010) 025011
  [arXiv:0811.1599 [hep-ph]].
  
\bibitem{Spalinski:1988az}
  M.~Spalinski,
  ``Chiral Corrections To The Axion Mass,''
  Z.\ Phys.\ C {\bf 41} (1988) 87.
  
\bibitem{Mao:2009sy}
  Y.~Y.~Mao {\it et al.} [TWQCD Collaboration],
  ``Topological Susceptibility to the One-Loop Order in Chiral Perturbation Theory,''
  Phys.\ Rev.\ D {\bf 80} (2009) 034502
  [arXiv:0903.2146 [hep-lat]].
  
\bibitem{Bijnens:2014lea}
  J.~Bijnens and G.~Ecker,
  ``Mesonic low-energy constants,''
  Ann.\ Rev.\ Nucl.\ Part.\ Sci.\  {\bf 64} (2014) 149
  [arXiv:1405.6488 [hep-ph]].

\bibitem{Aoki:2013ldr}
  S.~Aoki {\it et al.},
  ``Review of lattice results concerning low-energy particle physics,''
  Eur.\ Phys.\ J.\ C {\bf 74} (2014) 2890
  [arXiv:1310.8555 [hep-lat]].
  
\bibitem{Kaplan:1986ru}
  D.~B.~Kaplan and A.~V.~Manohar,
 ``Current Mass Ratios of the Light Quarks,''
  Phys.\ Rev.\ Lett.\  {\bf 56} (1986) 2004.
  


\bibitem{deDivitiis:2013xla}
  G.~M.~de Divitiis {\it et al.} [RM123 Collaboration],
  ``Leading isospin breaking effects on the lattice,''
  Phys.\ Rev.\ D {\bf 87} (2013) 11,  114505
  [arXiv:1303.4896 [hep-lat]].
  
  
\bibitem{Basak:2015lla}
  S.~Basak {\it et al.} [MILC Collaboration],
  ``Electromagnetic effects on the light hadron spectrum,''
  J.\ Phys.\ Conf.\ Ser.\  {\bf 640} (2015) 1,  012052
  [arXiv:1510.04997 [hep-lat]],
  and
  ``Electromagnetic effects on the light pseudoscalar mesons and determination of $m_u/m_d$,''
  talk by S.~Gotlieb at Lattice 2015, Japan, July 2015.

\bibitem{Horsley:2015eaa}
  R.~Horsley {\it et al.},
  ``Isospin splittings of meson and baryon masses from three-flavor lattice QCD + QED,''
  arXiv:1508.06401 [hep-lat].
  
\bibitem{Agashe:2014kda}
  K.~A.~Olive {\it et al.} [Particle Data Group Collaboration],
  ``Review of Particle Physics,''
  Chin.\ Phys.\ C {\bf 38} (2014) 090001.
  
\bibitem{Guo:2015oxa}
  F.~K.~Guo and U.~G.~Meißner,
  ``Cumulants of the QCD topological charge distribution,''
  Phys.\ Lett.\ B {\bf 749} (2015) 278
  [arXiv:1506.05487 [hep-ph]].
  
\bibitem{Bijnens:2001bb}
  J.~Bijnens, L.~Girlanda and P.~Talavera,
  ``The Anomalous chiral Lagrangian of order p**6,''
  Eur.\ Phys.\ J.\ C {\bf 23} (2002) 539
  [hep-ph/0110400].

\bibitem{Donoghue:1986wv}
  J.~F.~Donoghue, B.~R.~Holstein and Y.~C.~R.~Lin,
  ``Chiral Loops in pi0, eta0 ---> gamma gamma and eta eta-prime Mixing,''
  Phys.\ Rev.\ Lett.\  {\bf 55} (1985) 2766
   [Phys.\ Rev.\ Lett.\  {\bf 61} (1988) 1527].

  
\bibitem{Ananthanarayan:2002kj}
  B.~Ananthanarayan and B.~Moussallam,
  ``Electromagnetic corrections in the anomaly sector,''
  JHEP {\bf 0205} (2002) 052
  [hep-ph/0205232].

\bibitem{Giudice:2012zp}
  G.~F.~Giudice, R.~Rattazzi and A.~Strumia,
  ``Unificaxion,''
  Phys.\ Lett.\ B {\bf 715} (2012) 142
  [arXiv:1204.5465 [hep-ph]].
  
\bibitem{Kodaira:1979pa}
  J.~Kodaira,
  ``QCD Higher Order Effects in Polarized Electroproduction: Flavor Singlet Coefficient Functions,''
  Nucl.\ Phys.\ B {\bf 165} (1980) 129.

\bibitem{Jenkins:1990jv}
  E.~E.~Jenkins and A.~V.~Manohar,
 ``Baryon chiral perturbation theory using a heavy fermion Lagrangian,''
  Phys.\ Lett.\ B {\bf 255} (1991) 558.
  
 
 

\bibitem{QCDSF:2011aa}
  G.~S.~Bali {\it et al.} [QCDSF Collaboration],
  ``Strangeness Contribution to the Proton Spin from Lattice QCD,''
  Phys.\ Rev.\ Lett.\  {\bf 108} (2012) 222001
  [arXiv:1112.3354 [hep-lat]].

\bibitem{Engelhardt:2012gd}
  M.~Engelhardt,
  ``Strange quark contributions to nucleon mass and spin from lattice QCD,''
  Phys.\ Rev.\ D {\bf 86} (2012) 114510
  [arXiv:1210.0025 [hep-lat]].
 
\bibitem{Abdel-Rehim:2013wlz}
  A.~Abdel-Rehim, C.~Alexandrou, M.~Constantinou, V.~Drach, K.~Hadjiyiannakou, K.~Jansen, G.~Koutsou and A.~Vaquero,
  ``Disconnected quark loop contributions to nucleon observables in lattice QCD,''
  Phys.\ Rev.\ D {\bf 89} (2014) 3,  034501
  [arXiv:1310.6339 [hep-lat]].

\bibitem{Bhattacharya:2015gma}
  T.~Bhattacharya, R.~Gupta and B.~Yoon,
  ``Disconnected Quark Loop Contributions to Nucleon Structure,''
  PoS LATTICE {\bf 2014} (2014) 141
  [arXiv:1503.05975 [hep-lat]].
  
\bibitem{Abdel-Rehim:2015owa}
  A.~Abdel-Rehim {\it et al.},
  ``Nucleon and pion structure with lattice QCD simulations at physical value of the pion mass,''
  arXiv:1507.04936 [hep-lat].

\bibitem{Abdel-Rehim:2015lha}
  A.~Abdel-Rehim, C.~Alexandrou, M.~Constantinou, K.~Hadjiyiannakou, K.~Jansen, C.~Kallidonis, G.~Koutsou and A.~V.~Avilés-Casco,
  ``Disconnected quark loop contributions to nucleon observables using $N_f=2$ twisted clover fermions at the physical value of the light quark mass,''
  arXiv:1511.00433 [hep-lat].

\bibitem{Bhattacharya:2013ehc}
  T.~Bhattacharya, S.~D.~Cohen, R.~Gupta, A.~Joseph, H.~W.~Lin and B.~Yoon,
  ``Nucleon Charges and Electromagnetic Form Factors from 2+1+1-Flavor Lattice QCD,''
  Phys.\ Rev.\ D {\bf 89} (2014) 9,  094502
  [arXiv:1306.5435 [hep-lat]].
   
\bibitem{Yamanaka:2015lat}
N.~Yamanaka, H.~Ohki, S.~Hashimoto, T.~Kaneko (JLQCD Collab.), 
``Nucleon axial and tensor charges with the overlap fermions,''
talk presented at Lattice 2015 on 15-07-2015. 
 
\bibitem{Sikivie:2006ni}
  P.~Sikivie,
  ``Axion Cosmology,''
  Lect.\ Notes Phys.\  {\bf 741} (2008) 19
  [astro-ph/0610440].
 

\bibitem{Sikivie:1982qv}
  P.~Sikivie,
  ``Of Axions, Domain Walls and the Early Universe,''
  Phys.\ Rev.\ Lett.\  {\bf 48} (1982) 1156.
 
\bibitem{Vilenkin:1982ks}
  A.~Vilenkin and A.~E.~Everett,
  ``Cosmic Strings and Domain Walls in Models with Goldstone and PseudoGoldstone Bosons,''
  Phys.\ Rev.\ Lett.\  {\bf 48} (1982) 1867.

\bibitem{Vilenkin:1984ib}
  A.~Vilenkin,
  ``Cosmic Strings and Domain Walls,''
  Phys.\ Rept.\  {\bf 121} (1985) 263.

\bibitem{Davis:1986xc}
  R.~L.~Davis,
  ``Cosmic Axions from Cosmic Strings,''
  Phys.\ Lett.\ B {\bf 180} (1986) 225.

\bibitem{Bennett:1987vf}
  D.~P.~Bennett and F.~R.~Bouchet,
  ``Evidence for a Scaling Solution in Cosmic String Evolution,''
  Phys.\ Rev.\ Lett.\  {\bf 60} (1988) 257.
      
\bibitem{Dabholkar:1989ju}
  A.~Dabholkar and J.~M.~Quashnock,
  ``Pinning Down the Axion,''
  Nucl.\ Phys.\ B {\bf 333} (1990) 815.

\bibitem{Vincent:1996rb}
  G.~R.~Vincent, M.~Hindmarsh and M.~Sakellariadou,
  ``Scaling and small scale structure in cosmic string networks,''
  Phys.\ Rev.\ D {\bf 56} (1997) 637
  [astro-ph/9612135].

\bibitem{Kawasaki:2014sqa}
  M.~Kawasaki, K.~Saikawa and T.~Sekiguchi,
  ``Axion dark matter from topological defects,''
  Phys.\ Rev.\ D {\bf 91} (2015) 6,  065014
  [arXiv:1412.0789 [hep-ph]].
 

\bibitem{Berezhiani:1992rk}
  Z.~G.~Berezhiani, A.~S.~Sakharov and M.~Y.~Khlopov,
  ``Primordial background of cosmological axions,''
  Sov.\ J.\ Nucl.\ Phys.\  {\bf 55} (1992) 1063
   [Yad.\ Fiz.\  {\bf 55} (1992) 1918].

\bibitem{Masso:2002np}
  E.~Masso, F.~Rota and G.~Zsembinszki,
  ``On axion thermalization in the early universe,''
  Phys.\ Rev.\ D {\bf 66} (2002) 023004
  [hep-ph/0203221].

\bibitem{Graf:2010tv}
  P.~Graf and F.~D.~Steffen,
  ``Thermal axion production in the primordial quark-gluon plasma,''
  Phys.\ Rev.\ D {\bf 83} (2011) 075011
  [arXiv:1008.4528 [hep-ph]].
 
\bibitem{Salvio:2013iaa}
  A.~Salvio, A.~Strumia and W.~Xue,
  ``Thermal axion production,''
  JCAP {\bf 1401} (2014) 01,  011
  [arXiv:1310.6982 [hep-ph]].
 
 
\bibitem{Andersen:2011sf}
  J.~O.~Andersen, L.~E.~Leganger, M.~Strickland and N.~Su,
  ``Three-loop HTL QCD thermodynamics,''
  JHEP {\bf 1108} (2011) 053
  [arXiv:1103.2528 [hep-ph]].
 

 
\bibitem{Gasser:1986vb}
  J.~Gasser and H.~Leutwyler,
  ``Light Quarks at Low Temperatures,''
  Phys.\ Lett.\ B {\bf 184} (1987) 83.
 
\bibitem{Gasser:1987ah}
  J.~Gasser and H.~Leutwyler,
  ``Thermodynamics of Chiral Symmetry,''
  Phys.\ Lett.\ B {\bf 188} (1987) 477.

\bibitem{Hansen:1990yg}
  F.~C.~Hansen and H.~Leutwyler,
  ``Charge correlations and topological susceptibility in QCD,''
  Nucl.\ Phys.\ B {\bf 350} (1991) 201.
 
\bibitem{Gerber:1988tt}
  P.~Gerber and H.~Leutwyler,
  ``Hadrons Below the Chiral Phase Transition,''
  Nucl.\ Phys.\ B {\bf 321} (1989) 387.
 
\bibitem{Gross:1980br}
  D.~J.~Gross, R.~D.~Pisarski and L.~G.~Yaffe,
  ``QCD and Instantons at Finite Temperature,''
  Rev.\ Mod.\ Phys.\  {\bf 53} (1981) 43.

\bibitem{Linde:1980ts}
  A.~D.~Linde,
  ``Infrared Problem in Thermodynamics of the Yang-Mills Gas,''
  Phys.\ Lett.\ B {\bf 96} (1980) 289.

\bibitem{Rebhan:1993az}
  A.~K.~Rebhan,
  ``The NonAbelian Debye mass at next-to-leading order,''
  Phys.\ Rev.\ D {\bf 48} (1993) 3967
  [hep-ph/9308232].
  
\bibitem{Arnold:1995bh}
  P.~B.~Arnold and L.~G.~Yaffe,
  ``The NonAbelian Debye screening length beyond leading order,''
  Phys.\ Rev.\ D {\bf 52} (1995) 7208
  [hep-ph/9508280].
  
\bibitem{Kajantie:1997pd}
  K.~Kajantie, M.~Laine, J.~Peisa, A.~Rajantie, K.~Rummukainen and M.~E.~Shaposhnikov,
  ``Nonperturbative Debye mass in finite temperature QCD,''
  Phys.\ Rev.\ Lett.\  {\bf 79} (1997) 3130
  [hep-ph/9708207].
 
\bibitem{Philipsen:2000qv}
  O.~Philipsen,
  ``Debye screening in the QCD plasma,''
  hep-ph/0010327.
 
\bibitem{Maezawa:2007fc}
  Y.~Maezawa {\it et al.} [WHOT-QCD Collaboration],
  ``Heavy-quark free energy, debye mass, and spatial string tension at finite temperature in two flavor lattice QCD with Wilson quark action,''
  Phys.\ Rev.\ D {\bf 75} (2007) 074501
  [hep-lat/0702004].

\bibitem{Wantz:2009mi}
  O.~Wantz and E.~P.~S.~Shellard,
  ``The Topological susceptibility from grand canonical simulations in the interacting instanton liquid model: Chiral phase transition and axion mass,''
  Nucl.\ Phys.\ B {\bf 829} (2010) 110
  [arXiv:0908.0324 [hep-ph]].
   
\bibitem{Philipsen:2012nu}
  O.~Philipsen,
  ``The QCD equation of state from the lattice,''
  Prog.\ Part.\ Nucl.\ Phys.\  {\bf 70} (2013) 55
  [arXiv:1207.5999 [hep-lat]].
 
\bibitem{Borsanyi:2013bia}
  S.~Borsanyi, Z.~Fodor, C.~Hoelbling, S.~D.~Katz, S.~Krieg and K.~K.~Szabo,
  ``Full result for the QCD equation of state with 2+1 flavors,''
  Phys.\ Lett.\ B {\bf 730} (2014) 99
  [arXiv:1309.5258 [hep-lat]].
 
\bibitem{Ade:2015lrj}
  P.~A.~R.~Ade {\it et al.} [Planck Collaboration],
  ``Planck 2015 results. XX. Constraints on inflation,''
  arXiv:1502.02114 [astro-ph.CO].
 
\bibitem{Linde:1985yf}
  A.~D.~Linde,
  ``Generation of Isothermal Density Perturbations in the Inflationary Universe,''
  Phys.\ Lett.\ B {\bf 158} (1985) 375.
 
\bibitem{Hamann:2009yf}
  J.~Hamann, S.~Hannestad, G.~G.~Raffelt and Y.~Y.~Y.~Wong,
  ``Isocurvature forecast in the anthropic axion window,''
  JCAP {\bf 0906} (2009) 022
  [arXiv:0904.0647 [hep-ph]].
 
\bibitem{Sanfilippo:2015era}
  F.~Sanfilippo,
  ``Quark Masses from Lattice QCD,''
  PoS LATTICE {\bf 2014} (2015) 014
  [arXiv:1505.02794 [hep-lat]].

\bibitem{Mawhinney:lattice15}
R.~Mawhinney [RBC and UKQCD Collaborations], ``NLO and NNLO low energy constants
for SU(3) chiral perturbation theory,'' talk given at Lattice 2015, Japan, July 2015.  
 
\bibitem{Boyle:2015exm}
  P.~A.~Boyle {\it et al.},
  ``The Low Energy Constants of $SU(2)$ Partially Quenched Chiral Perturbation Theory from $N_{f}=2+1$ Domain Wall QCD,''
  arXiv:1511.01950 [hep-lat].
 
\bibitem{Altarelli:1988nr}
  G.~Altarelli and G.~G.~Ross,
  ``The Anomalous Gluon Contribution to Polarized Leptoproduction,''
  Phys.\ Lett.\ B {\bf 212} (1988) 391.
 
\bibitem{Larin:1993tq}
  S.~A.~Larin,
  ``The Renormalization of the axial anomaly in dimensional regularization,''
  Phys.\ Lett.\ B {\bf 303} (1993) 113
  [hep-ph/9302240].
 
 
\end{thebibliography}
\end{document}